\documentclass[apj]{emulateapj}
\usepackage{graphicx}
\usepackage{epstopdf}
\usepackage{longtable}
\usepackage{graphicx}
\usepackage{hyperref}
\usepackage{amsmath}
\usepackage{amssymb}	
\usepackage{aas_macros}
\usepackage{journals}
\usepackage{footnote}

\newcommand{\rion}[2]{{\ensuremath{\mbox{\rm #1$\,${\small\uppercase\expandafter{\romannumeral#2\relax}}}}}}

\shorttitle{Shocked Clouds in N132D}
\shortauthors{Dopita et al.}

\begin{document}

\title{Shocked Interstellar clouds and dust grain destruction\\ in the LMC Supernova Remnant N132D}

\author{Michael A. Dopita \altaffilmark{1, 2}, Fr\'ed\'eric P.A. Vogt  \altaffilmark{3, 4}, Ralph S. Sutherland \altaffilmark{1, 2}, \\ Ivo R. Seitenzahl \altaffilmark{5,6}, Ashley J. Ruiter \altaffilmark{5,6} \& Parviz Ghavamian \altaffilmark{7}}

\altaffiltext{1}{Research School of Astronomy and Astrophysics, Australian National University, Cotter Road, Weston Creek, ACT 2611, Australia.}
\altaffiltext{2}{ARC Centre of Excellence for All Sky Astrophysics in 3 Dimensions (ASTRO 3D).}
\altaffiltext{3}{European Southern Observatory, Av. Alonso de Cordova 3107, 763 0355 Vitacura, Santiago, Chile.}
\altaffiltext{4}{ESO Fellow.}
\altaffiltext{5}{School of Physical, Environmental and Mathematical Sciences, University of New South Wales, Australian Defence Force Academy, Canberra, ACT 2600, Australia.}
\altaffiltext{6}{ARC Future Fellow.}
\altaffiltext{7}{Department of Physics, Astronomy and Geosciences, Towson University, Towson, MD, 21252.}

\begin{abstract}
From integral field data we extract the optical spectra of 20 shocked clouds in the supernova remnant N132D in the Large Magellanic Cloud (LMC). Using self-consistent shock modelling, we derive the shock velocity, pre-shock cloud density and shock ram pressure in these clouds. We show that the [Fe X] and [Fe XIV] emission arises in faster, partially radiative shocks moving through the lower density gas near the periphery of the clouds. In these shocks dust has been effectively destroyed, while in the slower cloud shocks the dust destruction is incomplete until the recombination zone of the shock has been reached. These dense interstellar clouds provide a sampling of the general interstellar medium (ISM) of the LMC. Our shock analysis allows us to make a new determination of the ISM chemical composition in N, O, Ne, S, Cl and Ar, and to obtain accurate estimates of the fraction of refractory grains destroyed.  {From the derived cloud shock parameters, we estimate cloud masses and show that the clouds previously existed as typical self-gravitating Bonnor-Ebert spheres into which converging cloud shocks are now being driven.} 
\end{abstract}


\section{Introduction}\label{intro}
The supernova remnant (SNR) N132D in the Large Magellanic Cloud (LMC) is the brightest X-ray emitting SNR in this galaxy \citep{Hwang93, Favata97, Hughes98, Borkowski07, Xiao08}. Located within the stellar bar of the LMC,  N132D was first identified as a SNR by \citet{Westerlund66} on the basis of the association of a non-thermal radio source with a [\ion{S}{2}]-- bright optical structure. \citet{Danziger76} discovered high-velocity [\ion{O}{3}] and [\ion{Ne}{3}] material ejected { in the supernova explosion}, which established N132D as a relatively young SNR { in which the reverse shock has not yet reached the centre of the remnant, marking the formal transition to the Sedov phase of evolution. This would imply that the swept-up interstellar medium (ISM) is less than a few times the mass ejected in the supernova event, a result which is in apparent contradiction to the swept-up mass derived from X-ray observations \citep{Hughes98}.} From the dynamics of the fast-moving O and Ne material,  \citet{Danziger76} derived a maximum age of  3440\,yr for the SNR, and a probable age of $\sim 1350$\,yr. Subsequent analyses gave similar ages; 2350\,yr \citep{Lasker80}, and 2500\,yr \citep{Vogt11}.

Although most optical studies have concentrated on the fast-moving O- and Ne-rich ejecta \citep{Lasker80, Blair00, Vogt11}, there are a number of luminous, dense shock clouds with apparently `normal' interstellar composition and $\sim200$\,km/s velocity dispersion. The most prominent of these is the so-called \emph{Lasker's Bowl} structure in the northern part of the remnant, but HST imaging \citep{Blair00} reveals many other small complexes of shocked cloudlets. These clouds do not show appreciable enhancements in the strength of the [\ion{N}{2}] lines, and are thought to be simply ISM clouds over which the SNR blast wave has recently swept. In this paper we follow the study of N49 by \citet{Dopita16} in investigating the degree of dust destruction in these cloud shocks. We also derive the physical parameters for some 20 different shocked ISM cloudlets and estimate their chemical abundances. 

\section{Observations \& Data Reduction}\label{sec:obs}
The integral field spectra of N132D were obtained between 16 Nov 2017 and 24 Nov 2017 using the Wide Field Spectrograph (WiFeS)  \citep{Dopita07,Dopita10}, an integral field spectrograph mounted on the 2.3-m ANU telescope at Siding Spring Observatory. This instrument delivers a field of view of 25\arcsec $\times$ 38\arcsec at a spatial resolution of either 1.0\arcsec $\times$ 0.5\arcsec\ or 1.0\arcsec $\times$ 1.0\arcsec, depending on the binning on the CCD. In these observations, we operated in the binned 1.0\arcsec x 1.0\arcsec\ mode. The data were obtained in the low resolution mode $R \sim 3000$ (FWHM of $\sim 100$ km/s) using  the B3000 \& R3000 gratings in each arm of the spectrograph, with the RT560 dichroic which provides a transition between the  two arms at around 560nm. For details on the various instrument observing modes, see \citet{Dopita07}. 

All observations are made at PA=0, giving a long axis in the N-S direction. The basic grid consists of 15 pointings in an overlapping $3\times5$ grid (E-W:N-S), followed by 7 pointings centered on the overlap regions of the base $3\times5$  grid, with an additional pointing in the SE to probe the extent of the photoionised precursor region around N132D. The eighth overlap position was not observed due to deteriorating weather conditions. The typical seeing over the course of the observations was 1.5 arc sec. and ranged from 0.5 to 2.5 arc sec. over the individual exposures. Each region was observed with $2\times1000$\,s  exposure time giving a total integration time on target of 44000s. We used blind offsets from a reference star to move to each field of the mosaic consistently over the course of the observing run, and avoid any gaps in the resulting mosaic. A fault in the offset guide head gave rise to errors of $\lesssim 5$\,arc sec. in the pointing of the fields; small enough not to result in any gaps within the mosaic. 

The wavelength scale is calibrated using a series of Ne-Ar arc lamp exposures, taken throughout the night. Arc exposure times are 50s for the B3000 grating and 1s for the  R3000 grating. Flux calibration was performed using the STIS spectrophotometric standard stars  HD\,009051, HD\,031128 and HD\,075000 \footnote{Available at : \newline {\url{www.mso.anu.edu.au/~bessell/FTP/Bohlin2013/GO12813.html}}}. In addition, a B-type telluric standard HIP\,8352 was observed to better correct for the OH and H$_2$O telluric absorption features in the red. The separation of these features by molecular species allows for a more accurate telluric correction by accounting for night to night variations in the column density of these two species. All data cubes were reduced using the PyWiFeS \footnote {\url{http://www.mso.anu.edu.au/pywifes/doku.php.}} data reduction pipeline \citep{Childress14,Childress14b}. 

\begin{figure}
 \begin{centering}
  \includegraphics[scale = 0.4]{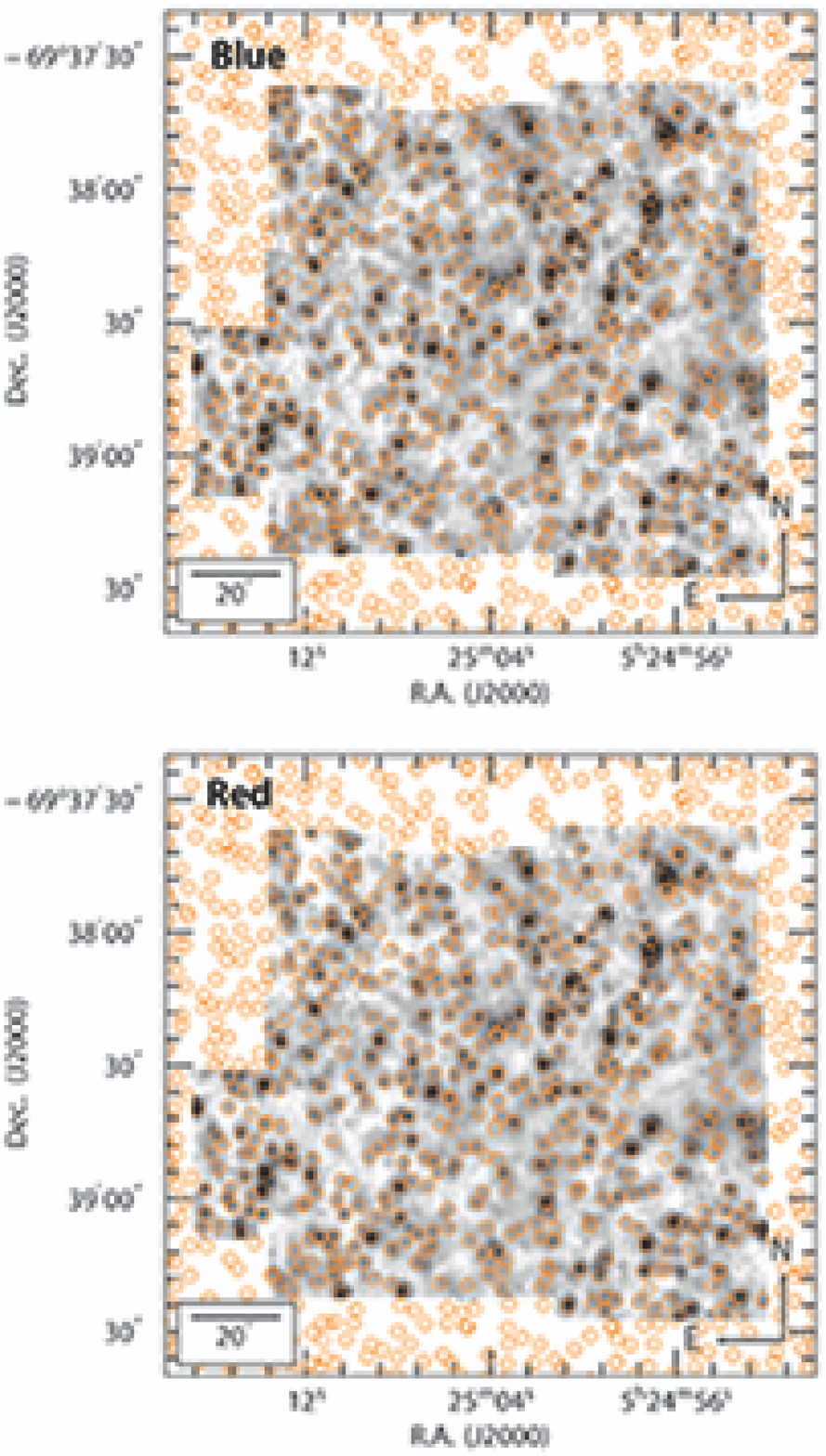}
  \end{centering}
  \caption{The alignment charts showing the reconstructed mosaic  in the summed 5600-5680\AA\ wavelength region for the blue arm (above) and the red arm (below). For reference, the positions of all the GAIA entries in the area are shown as orange circles.} \label{fig1}
 \end{figure}

All the individual, reduced cubes were median-averaged into a red and blue mosaic using a custom \textsc{python} script. The WCS solution of both mosaics is set by cross-matching their individual (collapsed) white-light images with all entries in the \textit{Gaia} \citep{GaiaCollaboration16a} DR1 \citep{GaiaCollaboration16}; see Fig \ref{fig1}). For simplicity, and given the spaxel size of the WiFeS datacubes, we restrict ourselves to integer shifts when combining fields. We estimate that the resulting absolute and relative (field-to-field) alignment accuracy are both $\pm$1\,arc sec.

\subsection{HST imaging of the cloudlets}
We have downloaded all the HST ACS F475W and F658N images from the \emph{Barbara A. Mikulski archive for space telescopes} (MAST).  The former filter encompasses the full velocity range of the [\ion{O}{3}]\,$\lambda\lambda$5959,5007\AA\ emission, whereas the latter covers the rest-frame H$\alpha$ emission. The dataset comprises four exposures of 380\,s each with the F475W filter, and four exposures of 360\,s each using the F658N filter, all acquired in January 2004 under program \#12001 (PI Green). 

We used the individual calibrated and CTE-corrected frames (\textsc{*\_flc.fits}) retrieved from MAST to construct a combined, drizzled image in both filters using the following steps. We first correct the WCS solutions of each frame using the \textsc{tweakreg} routine from the \textsc{drizzlepac 2.1.13} package via a dedicated \textsc{python} script. We anchor the WCS solution to the \textit{Gaia} \citep{GaiaCollaboration16a} DR1 \citep{GaiaCollaboration16} sources present within the images, retrieved automatically via the \textsc{astroquery} module. The WCS-corrected images are then merged using the \textsc{astrodrizzle} routine, with a pixel scale set to 0.04\,arcsec for both filters (for simplicity).

\begin{figure*}
  \includegraphics[width = \textwidth] {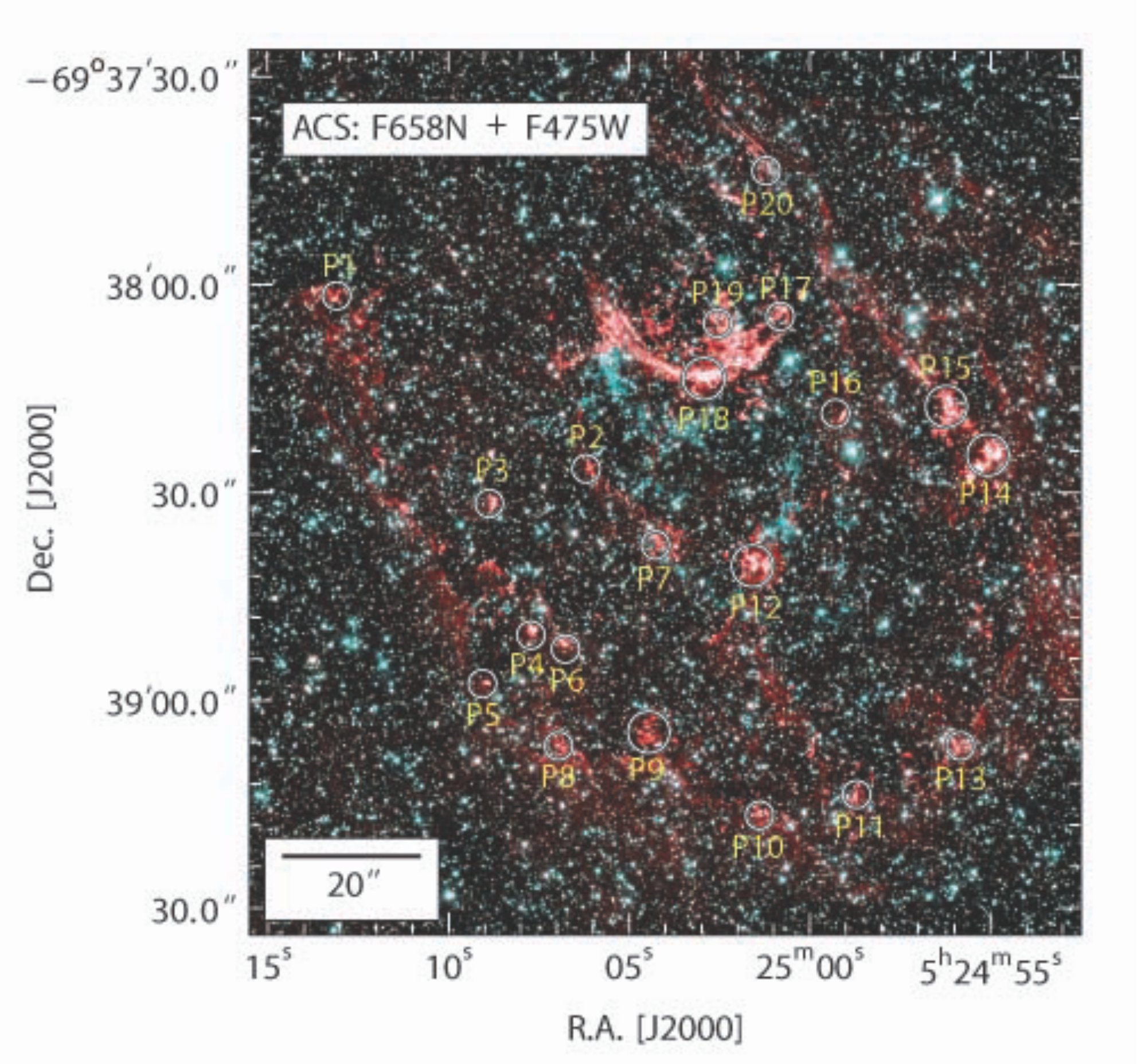}
  \caption{The shocked cloud regions listed in Table \ref{table1} are here displayed on an HST H$\alpha$ + [O III] image. The fast-moving  [O III] clouds appear as bright turquoise regions. Note that P20 is particularly strong in [O III] , due to the fact that the cloud shock here has not yet become fully radiative.
  } \label{fig2}
 \end{figure*}
 
 \begin{figure*}
 \includegraphics[width=\textwidth] {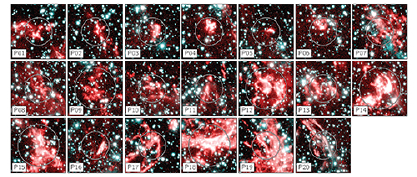}
  \caption{HST postage-stamp images of the 20 ISM clouds studied in this paper.  Each image covers  $8\times8$ arc sec. on the sky, or 1.78\,pc at the distance of the LMC. The extraction apertures used for the spectra are shown as circles.
  } \label{fig3}
 \end{figure*}

\subsection{Spectral Extraction}
The spectra of 20 prominent shocked ISM clouds were extracted from the global WiFeS mosaic datacubes using using {\tt QFitsView v3.1 rev.741}\footnote{{\tt QFitsView v3.1} is a FITS file viewer using the QT widget library and was developed at the Max Planck Institute for Extraterrestrial Physics by Thomas Ott.}.  The positions of the clouds were chosen so as to largely avoid contamination by high-velocity [{\ion{O}{3}] -- emitting material, { see Fig \ref{fig2}}. The exception is P07, but the velocity shift of the high velocity material here is sufficient as to allow accurate determinations of the narrow-line fluxes. We used a circular extraction aperture with a radius of either 2 or 3 arc sec. to best match the size of the bright region. To remove the residual night sky emission and (approximately) the faint stellar contribution, we subtracted a mean sky reference annulus 1 arc sec. wide surrounding the extraction aperture. The extraction regions were optimised by peaking up the signal in H$\alpha$ in the red data cube, and in H$\gamma$ in the blue data cube, respectively. This procedure may lead to some contamination by the faint extended cloud emission in the case of P07, P08, P14, P15, P18 and P20 (see Figure \ref{fig3}). However, the core region strongly dominates such fainter contributions. The positions of the extracted spectra, the extraction radius used, the measured mean radial velocity, and the H$\alpha$ velocity width (FWHM) (after correction for the instrumental resolution) of each cloud are listed in Table \ref{table1}, and the corresponding positions are shown on the HST ACS H$\alpha$ + [O III] image in Figure \ref{fig2} { and each cloud with its spectral extraction region is shown in an $8\times8$\,arc sec. postage stamp image in Figure \ref{fig3}}. A spectrum of the bright cloudlet P14 is shown in Figure \ref{fig4} to indicate the quality of the data. Note that the typical velocity FWHM of these clouds in only $\sim 180$\,km/s, which is fairly indicative of the mean shock velocity in those clouds \citep[Fig.6 in][]{Dopita12}. 

The mean Heliocentric radial velocity of all 20 clouds is +246\,km/s, and the mean Heliocentric radial velocity of the photoionised halo of N132D is found to be +261\,km/s, a difference which lies within the expected statistical error caused by sampling of the cloud shock velocities projected along the lines of sight within the random cloud positions inside the SNR. These figures can be compared to the +264.6\,km/s \citep{Smith71},  $+262\pm16$\,km/s \citep{Danziger76}, and +259\,km/s \citep{Feast64} for N127,  a nearby HII region.

Since most of the H$\alpha$ emission arises in the recombination zone of the cloud shocks, we can estimate a lower limit for the cloud shock velocities, $v_s$,  from the most extreme differences in the measured radial velocity of the cloud from the mean Heliocentric radial velocity of the photoionised halo of N132D. These are +173\,km/s for cloud P04, and -143\,km/s for cloud P19. Thus we have $v_s \gtrsim 160$\,km/s, in good agreement with the estimate based on the FWHM of the H$\alpha$ line profiles.

\begin{table}
 \centering
 \small
   \caption{The positions, extraction aperture sizes, and velocities of the shocked interstellar clouds of N132D}
    \label{table1}
   \scalebox{0.8}{
  \begin{tabular}{lccccc}
 \hline
 \hline
   Posn.& RA  & Dec & Ap.  Diam. & $V_{\mathrm {Hel.}}$ & $\Delta V_{\mathrm {FWHM}}$ \\
~\# & (J2000) &  (J2000) &   (arc sec.)  & $v_{\mathrm {Hel}}$ (km/s) & (km/s) \\
   \hline
P01 &  05:25:13.085  &  -69:38:01.6 & 4 & 312 & 190 \\
P02 &  05:25:06.188  &  -69:38:26.6 & 4 & 202 & 190 \\
P03 &  05:25:08.870  &  -69:38:31.6 & 4 & 273 & 203 \\
P04 &  05:25:07.720  &  -69:38:50.6 & 4 & 434 & 159 \\
P05 &  05:25:09.062  &  -69:38:57.6 & 4 & 256 & 216 \\
P06 &  05:25:06.762  &  -69:38:52.6 & 4 & 333 & 159 \\
P07 &  05:25:04.271  &  -69:38:37.6 & 4 & 204 & 201 \\
P08 &  05:25:06.953  &  -69:39:06.6 & 4 & 321 & 136 \\
P09 &  05:25:04.460  &  -69:39:04.6 & 6 & 259 & 155 \\
P10 &  05:25:01.391  &  -69:39:16.6 & 4 & 333 & 178 \\
P11 &  05:24:58.707  &  -69:39:13.6 & 4 & 373 & 173 \\
P12 &  05:25:01.587  &  -69:38:40.6 & 6 & 186 & 204 \\
P13 &  05:24:55.833  &  -69:39:06.6 & 4 & 321 & 150 \\
P14 &  05:24:55.074  &  -69:38:24.6 & 6 & 402 & 183 \\
P15 &  05:24:56.225  &  -69:38:17.6 & 6 & 321 & 198 \\
P16 &  05:24:59.291  &  -69:38:18.6 & 4 & 231 & 178 \\
P17 &  05:25:00.823  &  -69:38:04.6 & 4 & 214 & 158 \\
P18 &  05:25:02.932  &  -69:38:13.6 & 6 & 216 & 222 \\
P19 &  05:25:02.549  &  -69:38:05.6 & 4 & 118 & 152 \\
P20 &  05:25:01.211  &  -69:37:43.6 & 4 & 330 & 152 \\
\hline
 \end{tabular}}
\end{table}
 
 \begin{figure}
  \includegraphics[width = \columnwidth]{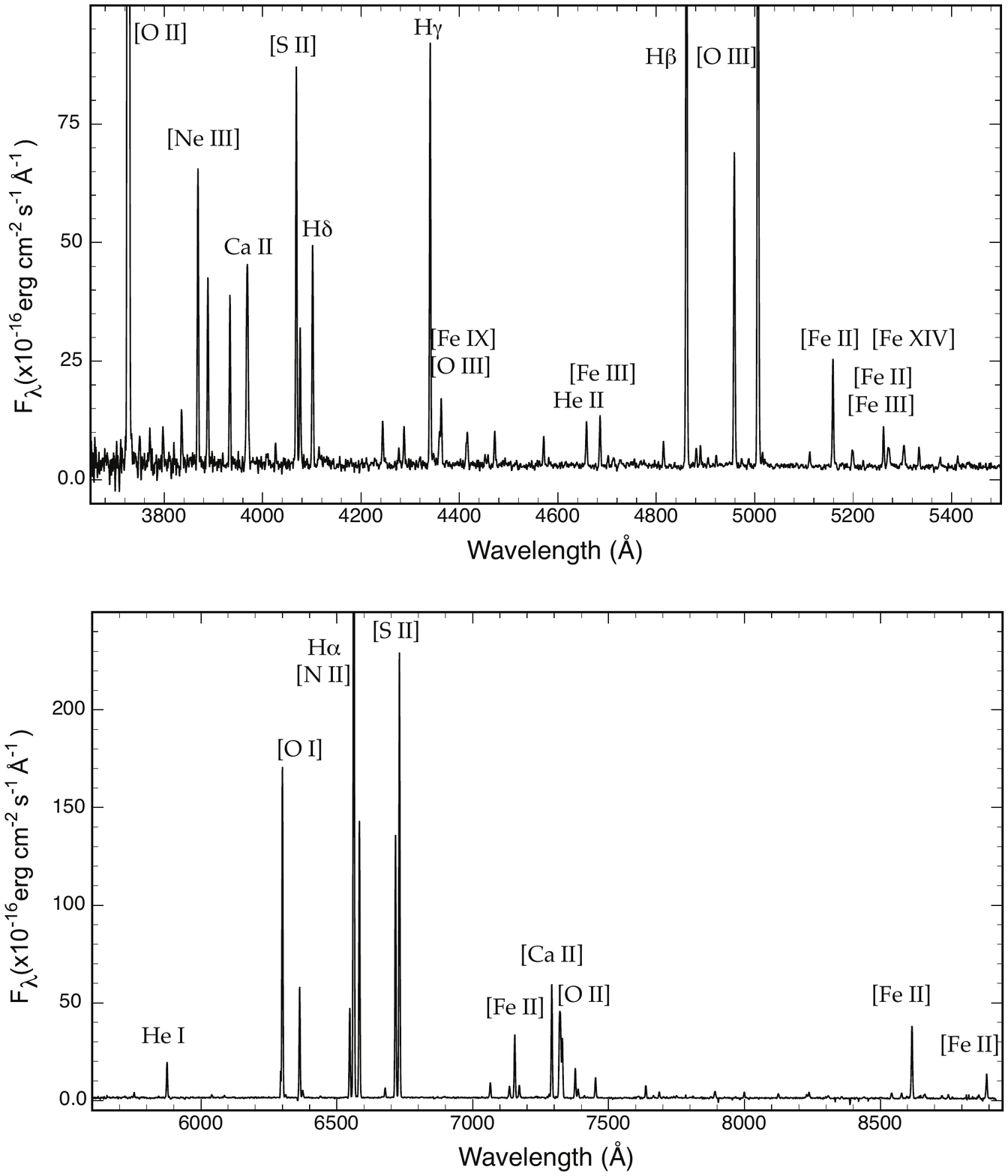}
  \caption{The extracted spectrum of the bright cloud P14 to show the quality of the data. Note the absence of both night sky emission lines, detectable atmospheric absorption, and underlying stellar continuum.The more prominent emission lines in the spectrum are identified according to ionic species.
  } \label{fig4}
 \end{figure}
 
\subsection{Measuring Emission Line Fluxes}
For each extracted spectrum, the spectra were first reduced to rest wavelength based on their measured radial velocities listed in Table \ref{table1}, and then emission-line fluxes { in units of erg/cm$^2$/s}, their uncertainties, the emission line FWHMs { (in \AA)} and the continuum levels were measured using the interactive routines in {\tt Graf} \footnote{Graf is written by R. S. Sutherland and is available at: {\url {https://miocene.anu.edu.au/graf}} }and in {\tt Lines} \footnote{Lines is written by R. S. Sutherland and is available at: {\url {https://miocene.anu.edu.au/lines}}}. The measured line fluxes { relative to H$\beta$} and their uncertainties { along with the absolute H$\beta$ fluxes in units of erg/cm$^2$/s}, are given for each of the clouds in Tables \ref{tableA1} -- \ref{tableA4} in the Appendix.

\section{Emission Line Diagnostics}\label{diagnostics}
\subsection{Self-Consistent Shock Modelling}
To analyse the spectrophotometry we have built a family of radiative shocks with self-consistent pre-ionisation using the MAPPINGS 5.12 code, following the methodology described in \citet{Sutherland17}, and applied to the study of Herbig-Haro objects by \citet{Dopita17b}. The pre-shock density is set by the ram pressure of the shock, taken to be independent of shock velocity; $P_{\mathrm{ram}} = 1.5\times 10^{-7}$\,dynes cm$^{-2}$, to ensure that the measured [\ion{S}{2}] densities  (which give the electron density near the recombination zone of the shock) approximately match those produced by the models. For each abundance set, a set of models was run for $100 \leqslant v_s \leqslant 475$\,km/s in steps of 25\,km/s. The magnetic field pressure in the un-shocked cloud ahead of the photo-ionised precursor is assumed to be in equipartition with the gas pressure, $P_{\mathrm{mag}} = P_{\mathrm{gas}}$, and the temperature of the gas entering the shock is given by the self-consistent pre-ionisation computation; see \citet{Sutherland17} for details.

The abundance set was initially taken as being 0.5 times the Local Galactic Concordance (LGC) values \citep{Nicholls17}, but was then iterated manually to achieve a better fit of theory to the spectra of the clouds on the standard \citet{BPT81} and \citet{Veilleux87} diagnostic diagrams. The depletion factors of the heavy elements caused by the condensation of these elements onto dust are defined as the ratio of the gas phase abundance to the total element abundance. The  depletion factors are derived from the formulae of \citet{Jenkins09}, extended to the other elements on the basis of their condensation temperatures and/or their position on the periodic table. In these  shock models, following \citet{Dopita16} we have investigated the effect of changing the logarithmic Fe depletion,  $\log D_{\rm Fe}$ in the range $0.0 > \log \left [ D_{\rm Fe} \right ] > -1.0$. 

The abundance set adopted for the theoretical grid at the various values of $\log D_{\rm Fe}$ is given in Table \ref{table2}. The abundances of C, Mg, and Si are not constrained by our observations and for these we fix the abundances at half of the LGC values. We produce a more refined estimate of the LMC chemical abundances when we build detailed models for the brightest clouds, below.

\begin{table}
 \caption{The LMC abundance set used in the grid, and the corresponding gas phase abundances for various logarithmic depletion factors  in Fe.}\label{table2}
 \scalebox{1.0}{
\begin{tabular}{lccccc}
 \hline
 \hline
        & \multicolumn{5}{|c|}{$12 + \rm {\log(X/H)}$}\\\cline{2-6}
        &                  &     \\
Element &   Total   &   $\log D_{\rm Fe}$  & $\log D_{\rm Fe}$  & $\log D_{\rm Fe}$  & $\log D_{\rm Fe}$  \\
 &   Abund.  &    -0.25   &  -0.50  &  -0.75 &  -1.00\\

\hline
  H     &   12.00  &   12.00  &  12.00 & 12.00 & 12.00\\               
  He    &   10.96      &   10.96   & 10.96 & 10.96 & 10.96 \\      
  C     &   8.01   &   7.99  & 7.97 & 7.95 &  7.93 \\             
  N     &   7.30   &   7.30   & 7.30 & 7.30 & 7.30\\            
  O     &   8.25   &   8.25   & 8.25 & 8.24 & 8.23 \\             
  Ne    &   7.79   &   7.79  & 7.79 & 7.79 & 7.79 \\             
  Mg    &   7.26    &   7.26 &  7.25 &  7.15 & 6.95 \\              
  Si    &   7.20   &   7.20   & 7.20 &  7.15 & 6.93 \\            
  S     &   7.00  &   7.00   & 7.00 & 7.00 & 7.00 \\             
  Cl    &   5.20    &   5.20 & 5.20 & 5.20 & 5.20 \\              
  Ar    &   5.90   &   5.90   & 5.90 & 5.90 & 5.90 \\            
  Ca    &   6.02   &   6.02   & 5.66 &  5.27 & 4.87 \\
  Fe    &   7.22   &   6.97   & 6.72 &  6.47 & 6.22 \\             
  Ni    &   5.90    &   5.78   &  5.49 &  5.19 & 4.91\\
\hline
\end{tabular}} \\
\end{table}

The line intensities and line ratios given in this paper comprise the sum of the radiative shock and its photoionised precursor. In the case of the fastest shock, the extent of the photoionised precursor region and/or the cooling length of the shock may exceed the physical extent of the pre-shocked cloud. In Figure \ref{fig5} we show the cooling length of the shock to 1000K, the depth of the photoionised precursor to the point where hydrogen is only 1\% ionised, and the fraction of the shock H$\beta$ emission which arises from the precursor. The cooling length and precursor length are given for a ram-pressure of $P_{\mathrm{ram}} = 1.0\times 10^{-7}$\,dynes cm$^{-2}$. The shock cooling length scales inversely as the ram pressure, and the precursor length remains approximately constant. At the distance of the LMC 0.1\,pc corresponds to 0.45\,arc sec.

\begin{figure}
  \includegraphics[scale=0.4]{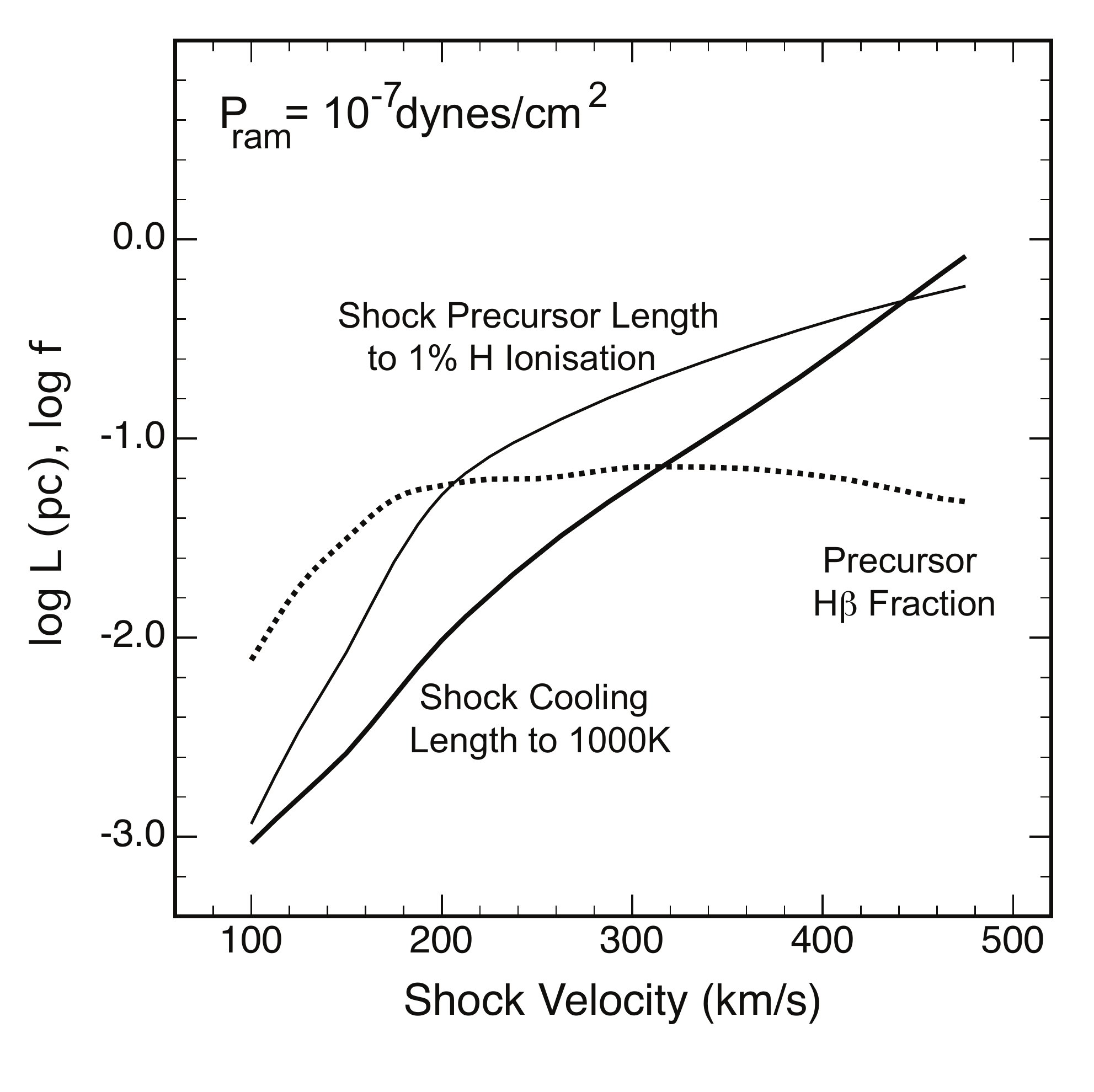}
  \caption{The computed cooling length of the shock to $T_e = 1000$\,K (thick line), the thickness of the photoionised precursor to the point where hydrogen is only 1\% ionised (thin line), and the fraction of the shock H$\beta$ emission which arises from the precursor (dotted line). These are all given for a ram pressure of  $P_{\mathrm{ram}} = 1.0\times 10^{-7}$\,dynes cm$^{-2}$. At the densities of those models, the shock cooling length scales inversely as the ram pressure, while the precursor length is almost unaffected by changes in the ram pressure. 
  } \label{fig5}
 \end{figure}

\subsection{Shock Velocity Diagnostics}
Here we investigate potentially useful diagnostics of the shock velocity using emission line ratios. To avoid issues of chemical abundance difference, the line ratio used should involve different levels of the same ion, or else different ions of the same element. In \citet{Dopita16}, we already noted that the [\ion{O}{3}] $\lambda\lambda 4363/5007$ ratio was a useful indicator of shock velocity. This is because the [\ion{O}{3}] temperature in the cooling zone of the shock is high at low shock velocities (faster than $\sim85$\,km/s, when this ion is first produced in the cooling zone of the shock), reaching a maximum at $v_s \sim 140$\,km/s when \ion{O}{3} starts to become ionised to higher ionisation stages. As the velocity increases further, \ion{O}{3} is ionised to \ion{O}{4}, and the  [\ion{O}{3}] emission becomes confined to the cooler region nearer the recombination zone of the shock. At still higher velocities, a  \ion{O}{3} zone photoionised by EUV photons generated in the  cooling zone of the shock develops adjacent to the recombination zone. This has still lower mean temperature; $T_e\lesssim 10^4$\,K. For these physical reasons, the [\ion{O}{3}] temperature is more or less a decreasing function of shock velocity, and is relatively insensitive to  $\log D_{\rm Fe}$. This behaviour is shown in Figure \ref{fig6}.

A second diagnostic is the excitation of He, as measured by the \ion{He}{2}/\ion{He}{1}  $\lambda\lambda 4686/5876$ ratio. This is shown in the second panel of Figure \ref{fig6}. Below $v_s \sim 200$\,km/s, this ratio is determined by the pre-ionisation of the shock, and the temperature structure of the post-shock region, and is multi-valued. However, above this velocity, it provides a useful diagnostic, albeit somewhat sensitive to $\log D_{\rm Fe}$. This sensitivity is caused by changes in the cooling length with gas-phase Fe abundance, which changes the ionisation structure of He in the shock. In what follows, we will use both the He and [\ion{O}{3}] diagnostic ratios as spectroscopic indicators of the shock velocity.

\begin{figure}
  \includegraphics[width = \columnwidth]{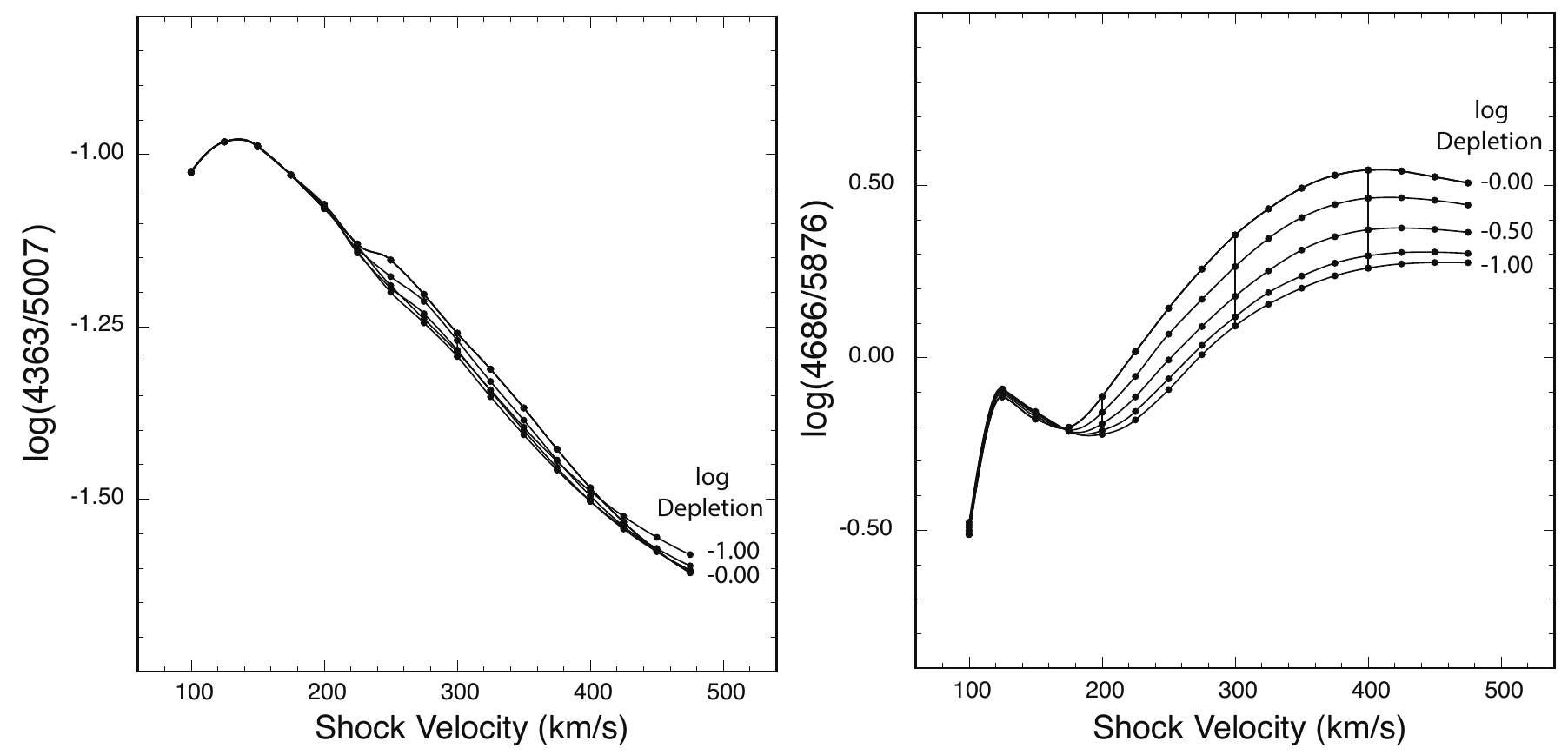}
  \caption{The shock velocity sensitive line ratios  [\ion{O}{3}] $\lambda\lambda 4363/5007$ (left) and \ion{He}{2}/\ion{He}{1}  $\lambda\lambda 4686/5876$ (right) as a function of shock velocity and iron depletion factor; $\log D_{\rm Fe}$.
  } \label{fig6}
 \end{figure}
  \begin{figure}
  \includegraphics[width = \columnwidth]{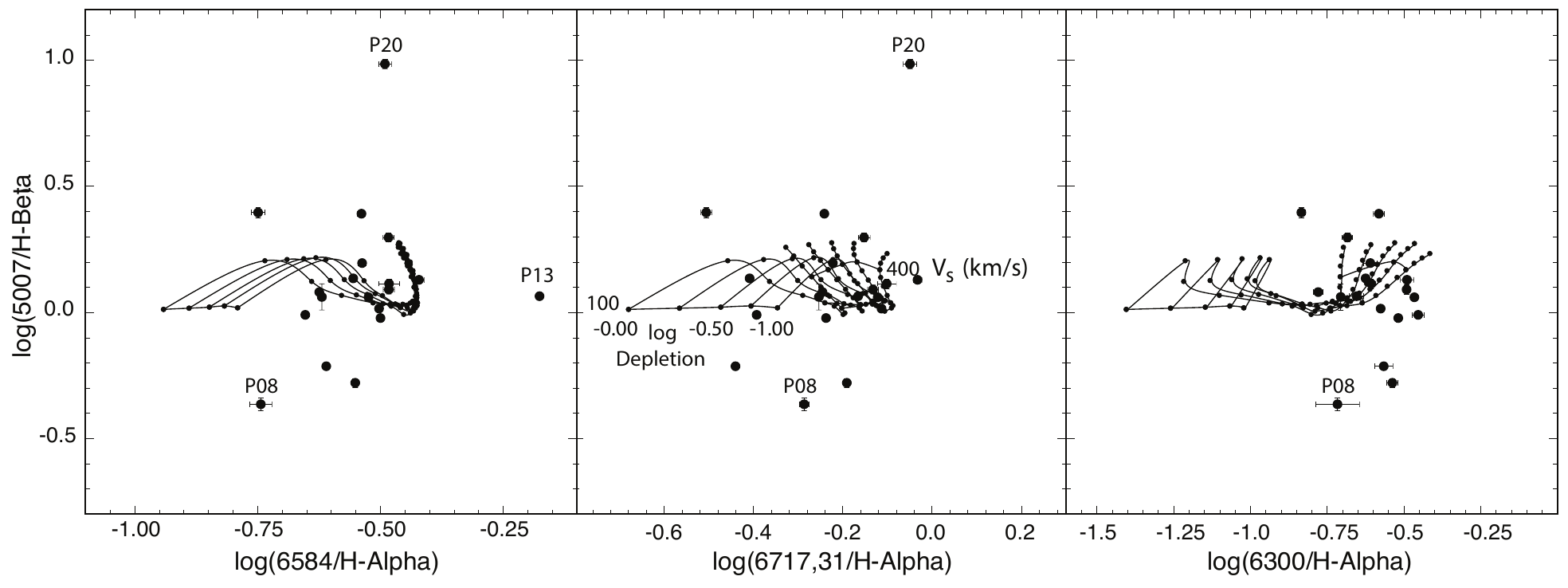}
  \caption{The standard BPT plots showing the positions of our clouds superimposed on the theoretical grid. Overall, the model grid reproduces the observations in a satisfactory manner. Anomalous regions are indicated. These are discussed further in Section \ref{cloudshocks}.
  } \label{fig7}
 \end{figure}

\subsection{BPT diagnostic diagrams}
For shocks, the position of the observations on the standard \citet{BPT81} and \citet{Veilleux87} diagnostic diagrams (hereafter BPT Diagrams), is indicative of the mode of excitation. However, they are not really useful as diagnostics of the detailed shock conditions. This is demonstrated in Figure \ref{fig7}. It should be noted that the range in observed line ratio and in the theoretical grid is very small on these diagrams, compared with the range covered by either HII regions or active galaxies and Narrow Line Regions. All these diagrams serve to show is that the theoretical models and the observations show a satisfactory overlap. 

This said, a few points on these diagrams are worthy of further mention. First P13 is the only region which shows an excess in the line intensity of [\ion{N}{2}], which may be indicative of pre-supernova mass-loss enrichment. However, this seems a little unlikey as the region is close to the boundary of the SNR shell. Second, P20 shows extremely strong [\ion{O}{3}] emission. This is evidenced by the blue appearance of this region in Figure \ref{fig3}. This region will be discussed in more detail below, where we show that it is due to a finite-age shock in which the gas has not had time to become fully radiative. Lastly, P08 shows particularly weak  [\ion{O}{3}] emission. This region, also discussed below, lies in a complex of Balmer dominated or non-radiative filaments in which a high velocity non-radiative shock is passing through an un-ionised or partially-ionised precursor medium.

 \subsection{ [\ion{Fe}{2}], [\ion{Fe}{3}] and  [\ion{Fe}{7}] diagnostics}
As can be seen from Tables \ref{tableA1} - \ref{tableA4} the [\ion{Fe}{2}] spectrum is very rich. For the purposes of analysis, we have selected the brightest of these lines, [\ion{Fe}{2}]${\lambda 5158}$ and  [\ion{Fe}{2}]${\lambda 8617}$. In Figure \ref{fig8} we present the observed points superimposed on our theoretical shock model grid. From this, it is immediately apparent that nowhere does the measured  iron depletion factor; $\log D_{\rm Fe}$ exceed -0.5. Indeed many of the points are consistent with no depletion at all. 

  \begin{figure}
  \includegraphics[width = \columnwidth]{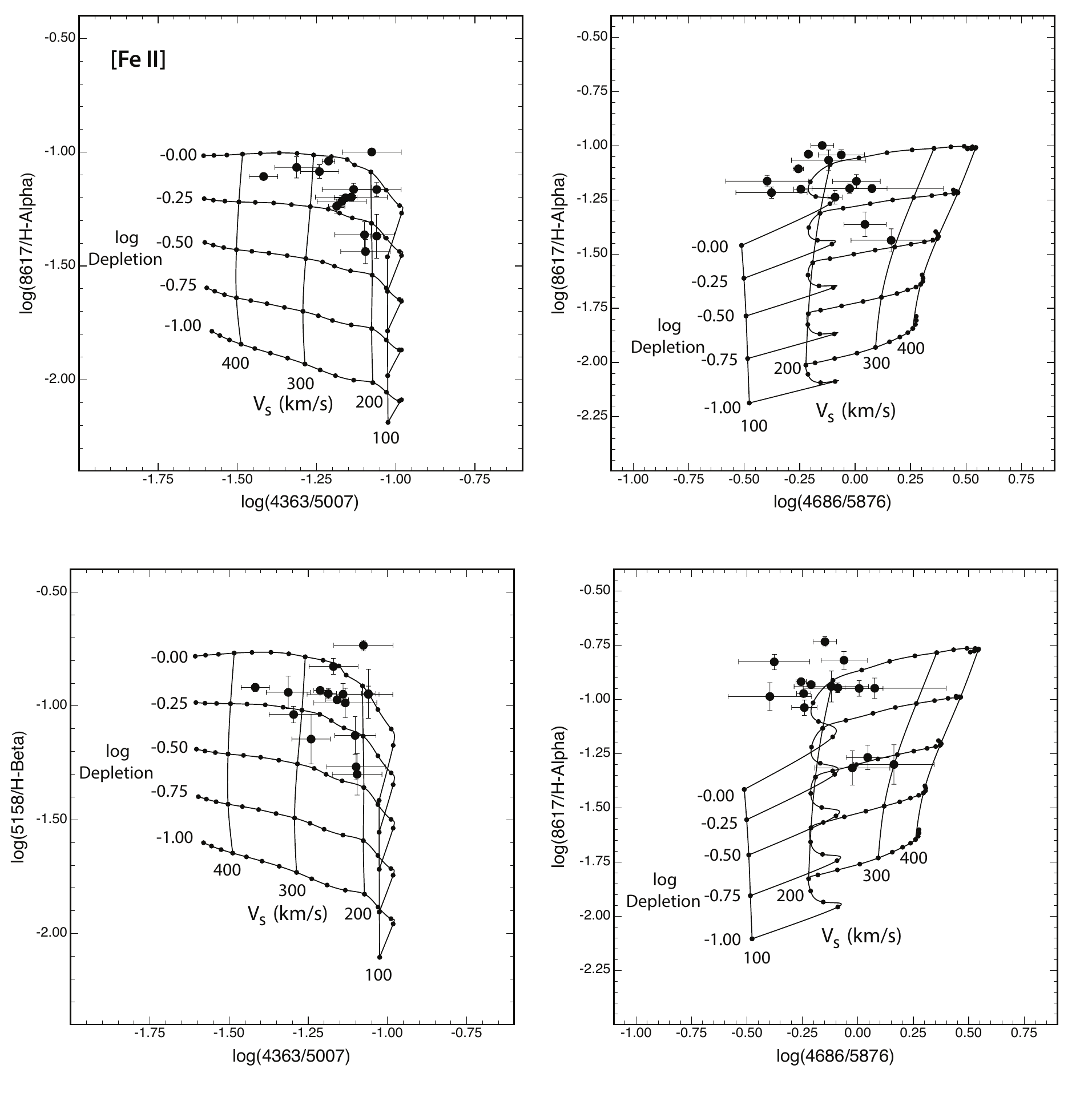}
  \caption{The depletion of iron as measured by the strong [\ion{Fe}{2}] lines. The data are consistent with a mean depletion factor $\log D_{\rm Fe} \sim -0.14$
  } \label{fig8}
 \end{figure}
 
\begin{figure}
  \includegraphics[width = \columnwidth]{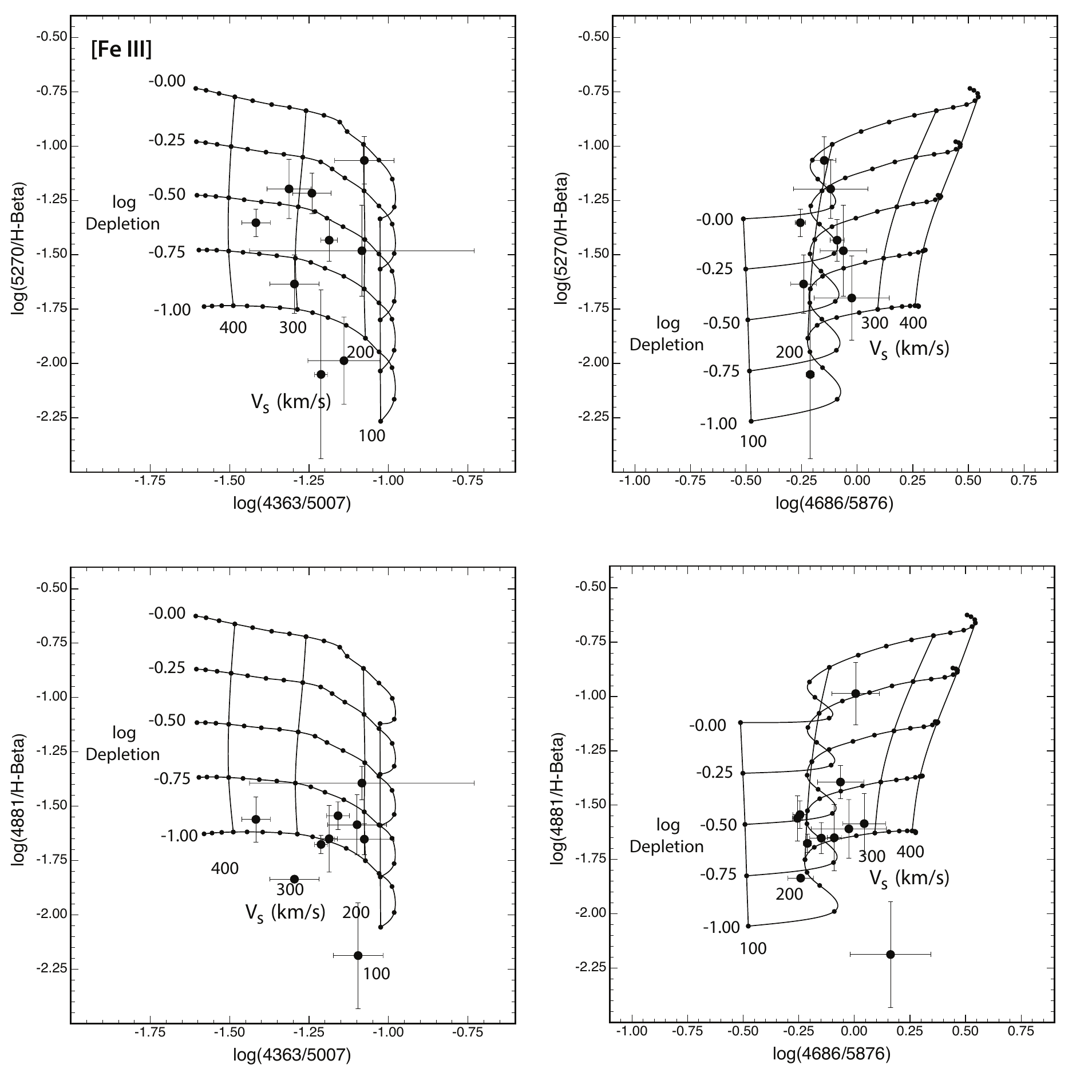}
  \caption{The depletion of iron using the strong [\ion{Fe}{3}] lines. The data are consistent with a mean depletion factor $\log D_{\rm Fe} \sim -0.74$
  } \label{fig9}
 \end{figure}

 \begin{figure}
  \includegraphics[width = \columnwidth]{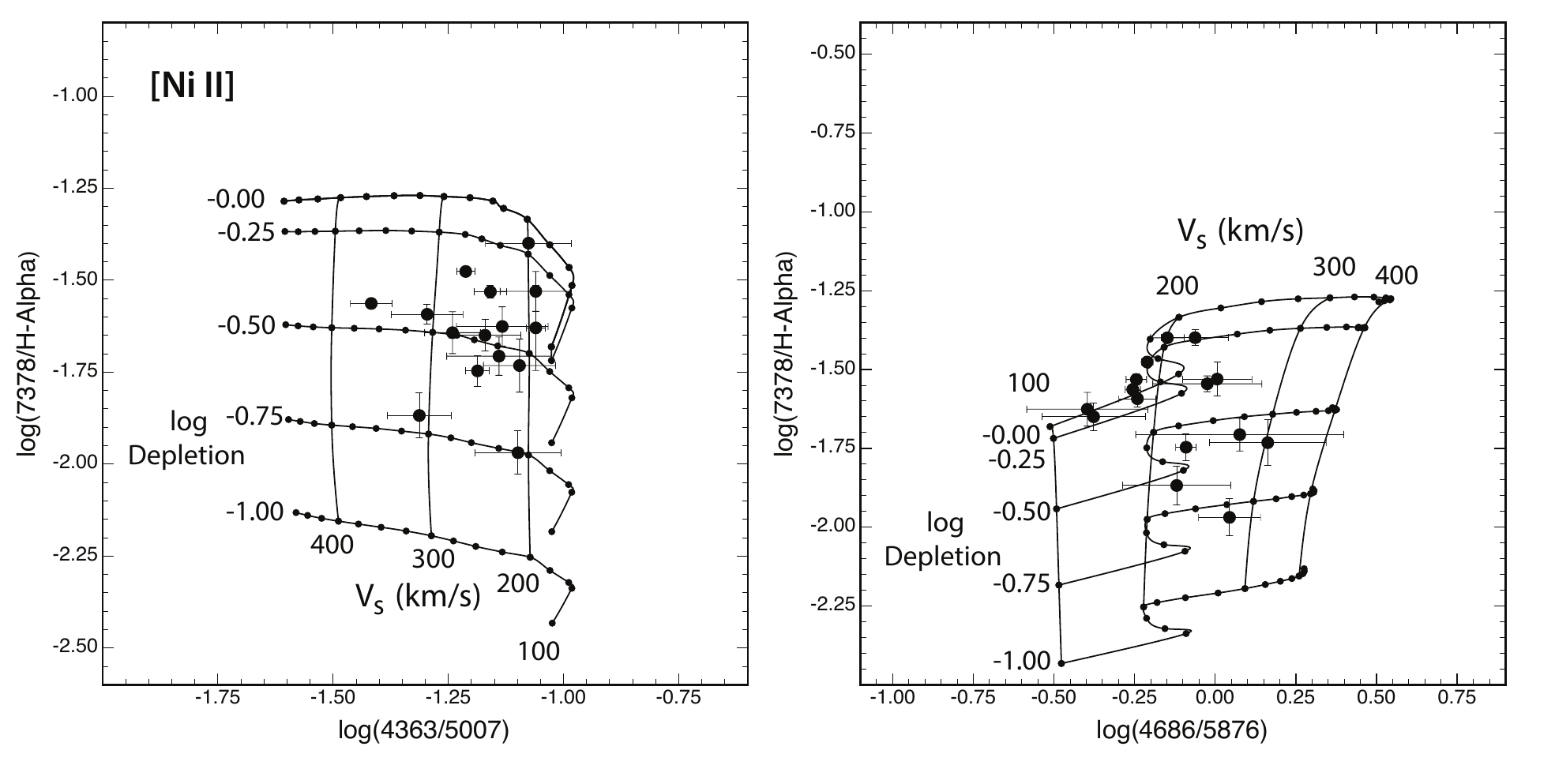}
  \caption{The depletion of Fe using the  [\ion{Fe}{7}]$\lambda 6087$/H$\alpha$ ratio. The data are consistent with a mean depletion factor $\log D_{\rm Fe} \sim -0.4$ and a range of $\sim -0.0 - -0.75$.
  } \label{fig11}
 \end{figure}

\begin{figure}
  \includegraphics[width = \columnwidth]{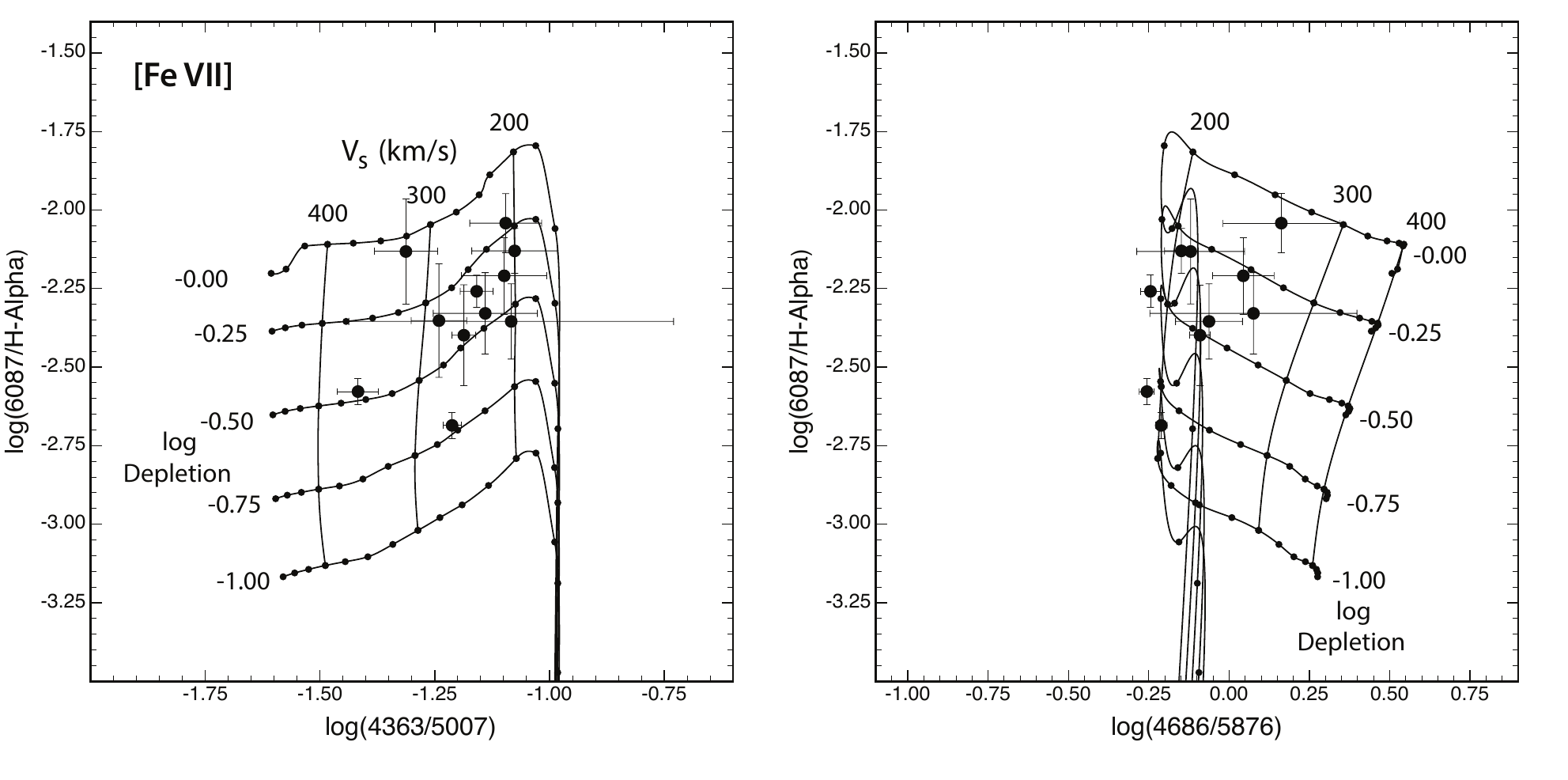}
  \caption{The depletion of Ni using the  [\ion{Ni}{2}]$\lambda 7378$/H$\alpha$ ratio. The data are consistent with a mean depletion factor $\log D_{\rm Fe} \sim -0.4$ and a range of $\sim -0.0 - -0.75$.
  } \label{fig10}
 \end{figure}
 
 \begin{figure}
  \includegraphics[width = \columnwidth]{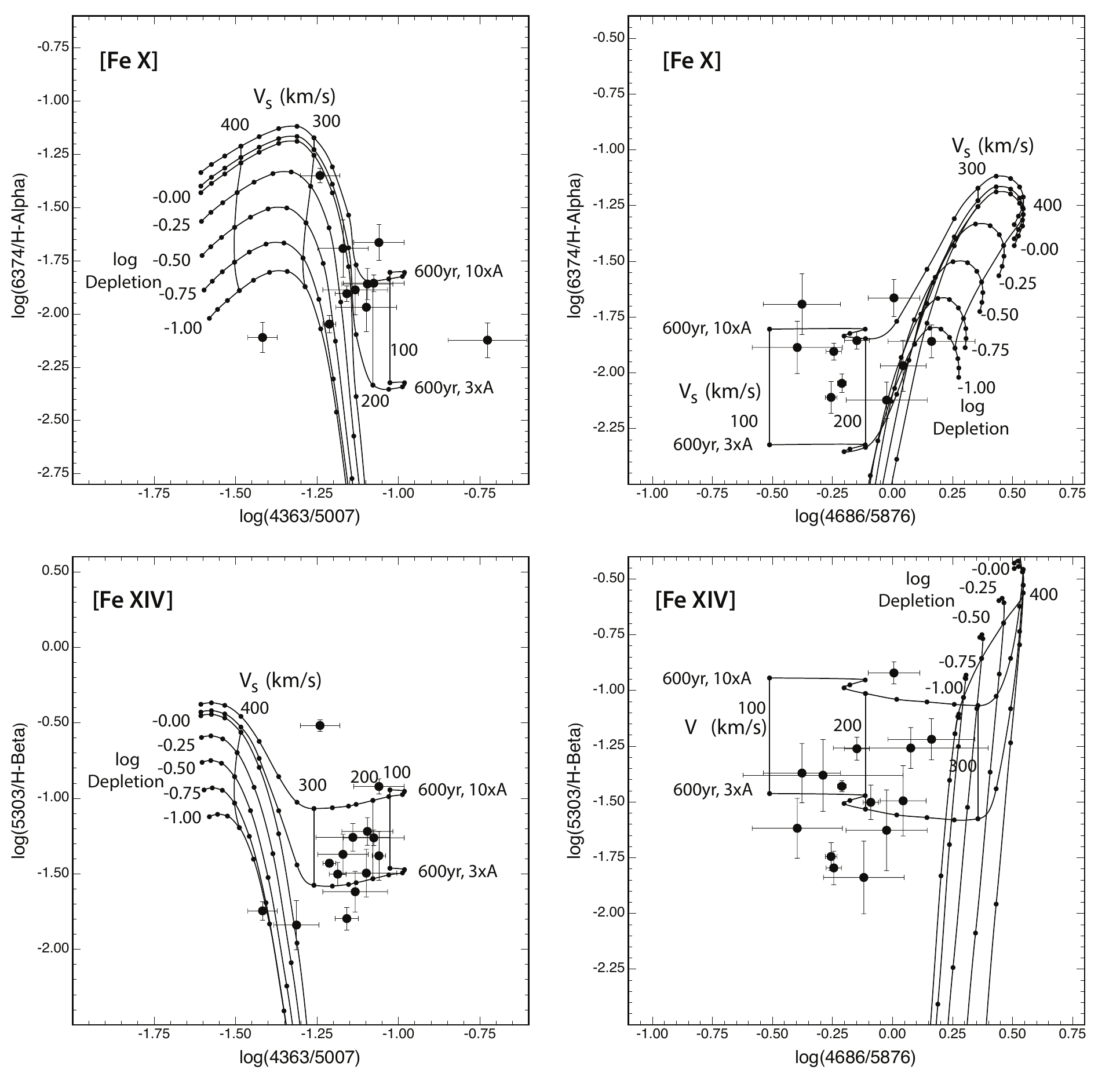}
  \caption{The [\ion{Fe}{10}]$\lambda 6374$/H$\alpha$ and [\ion{Fe}{14}]$\lambda 5303$/H$\beta$. The standard shock grid, as presented in the previous figures clearly fails to reproduce the positions of the observed points on these diagnostic diagrams. However, by adding a faster, partially radiative shock of age $\sim 600$\,yr, which covers an area of 3-10 times the area of the dense cloud shock, we can achieve a satisfactory fit with the observations (upper two curves marked 600yr, 3xA and 600yr, 10xA, respectively).
  } \label{fig12}
 \end{figure}

 \begin{table}
 \centering
 \small
   \caption{Shock velocities derived from the [\ion{O}{3}] line ratio and estimated \ion{Fe}{2} depletion factors of the shocked interstellar clouds of N132D}
    \label{table3}
   \scalebox{0.9}{
  \begin{tabular}{lccccc}
 \hline
 \hline
 Line:  &  & \bf{[Fe II] }  &  & \bf{[Fe II] } & \\
Posn.  & $V_{\mathrm{OIII}}$ & $\bf{\lambda 5158}$  &  &  $\bf{\lambda 8617}$ & \\
  \# & (km/s) & $\log D_{\rm Fe}$ & Err. & $\log D_{\rm Fe}$ & Err. \\
 \hline
 P02 & $190\pm 40$ & 0.00 & 0.03 & -0.05 & 0.02\\
P03 & $225\pm 50$ & -0.13 & 0.07 & -0.12 & 0.03\\
P04 & $200\pm  50$ & 0.20 & 0.05 & 0.10 & 0.02\\
P05 & $325\pm 35$ & -0.19 & 0.06 & -0.09 & 0.05\\
P06 & ...   & 0.00 & 0.05 & 0.00 & 0.05\\
P07 & $230\pm 70$ & -0.10 & 0.05 & -0.15 & 0.02\\
P10 & $210\pm 40$ & -0.42 & 0.10 & -0.38 & 0.05\\
P11 & $285\pm  35$ & -0.32 & 0.12 & -0.10 & 0.03\\
P12 & $260\pm  25$ & -0.13 & 0.02 & -0.23 & 0.03\\
P13 & $250\pm  40$ & -0.02 & 0.03 & -0.20 & 0.03\\
P14 & $280\pm 20$ & -0.11 & 0.02 & -0.02 & 0.02\\
P15 & $240\pm 25$ & -0.13 & 0.02 & -0.15 & 0.02\\
P16 & $210\pm 50$ & -0.40 & 0.06 & -0.32 & 0.07\\
P17 & $210\pm 40$ & -0.25 & 0.09 & É & \\
P18 & $360\pm 20$ & -0.15 & 0.02 & -0.12 & 0.02\\
P19 & $310\pm 40$ & -0.27 & 0.04 & É & \\
P20 & $190\pm 20$ & 0.00 & 0.12 & -0.30 & 0.10\\
 &    &  &  &  & \\
 \hline
Av: & 248  & -0.14 &  & -0.14 & \\
\hline
 \end{tabular}}
\end{table}
\begin{table}
 \centering
 \small
   \caption{Shock velocities derived from the \ion{He}{2}/\ion{He}{1} line ratio and estimated \ion{Fe}{2} depletion factors of the shocked interstellar clouds of N132D}
    \label{table4}
   \scalebox{0.9}{
  \begin{tabular}{lccccc}
  \hline
  \hline
 Line:  &  & \bf{[Fe II] }  &  & \bf{[Fe II] } & \\
Posn.  & $V_{\mathrm{He}}$ & $\bf{\lambda 5158}$  &  &  $\bf{\lambda 8617}$ & \\
  \# & (km/s) & $\log D_{\rm Fe}$ & Err. & $\log D_{\rm Fe}$ & Err. \\
 \hline
P01 & $225\pm55$ & -0.50 & 0.10 & -0.15 & 0.03\\
P02 & $225\pm 25$ & -0.10 & 0.04 & -0.10 & 0.03\\
P03 & ... & ... &  & -0.10 & 0.05\\
P04 & $200\pm 25$ & 0.20 & 0.05 & 0.10 & 0.02\\
P05 & $200\pm 50$ & -0.03 & 0.06 & 0.02 & 0.02\\
P06 & $210\pm 24$ & 0.08 & 0.05 & 0.03 & 0.03\\
P07 & $245\pm 100$ & -0.12 & 0.05 & -0.20 & 0.02\\
P10 & $300\pm 75$ & -0.55 & 0.15 & -0.48 & 0.05\\
P12 & $210\pm  20$ & -0.07 & 0.02 & -0.20 & 0.04\\
P14 & $175\pm  15$ & 0.08 & 0.02 & 0.12 & 0.02\\
P15 & $170\pm  20$ & 0.05 & 0.02 & -0.10 & 0.02\\
P16 & $265\pm 30$ & -0.49 & 0.08 & -0.35 & 0.06\\
P18 & É   &  &  & 0.00 & 0.03\\
P20 & É   &  &  & -0.27 & 0.10\\
 &  &  &  &   & \\
 \hline
Av: & 220   & -0.13 &  & -0.12 & \\
\hline
 \end{tabular}}
\end{table}

Furthermore, it is clear that the indicated shock velocities in general exceed 200\,km/s. Thus for most of these the  [\ion{O}{3}] line ratio can be used  to estimate the shock velocity. However, a number of the points lie in the ambiguous region of the  \ion{He}{2}/\ion{He}{1} line ratio velocity diagnostic. To the extent to which it is possible, both shock velocities and $\log D_{\rm Fe}$ have been estimated for each cloud from these diagnostic diagrams. The results are given in Tables \ref{table3} and \ref{table4}. We find a mean shock velocity $\left< v_s \right >$ of 240 km/s, and  $\left < \log D_{\rm Fe} \right > = -0.15$. This result agrees very closely with the $\left < \log D_{\rm Fe} \right > = -0.16\pm0.07$ found by \citet{Dopita16} in N49, another bright LMC SNR, and shows that dust has been largely destroyed by the time the recombination zone of the shock has been reached.

A similar analysis was applied to the [\ion{Fe}{3}]$\lambda 5270$/H$\beta$ and  [\ion{Fe}{3}]$\lambda 4881$/H$\beta$ line ratios. This is shown in Figure \ref{fig9}. The two lines agree less well with each other, the [\ion{Fe}{3}]$\lambda 4881$/H$\beta$ ratio tending to produce a larger depletion factor. In each line ratio, a much wider range of depletion factors is observed, from -0.25 to $\sim -1.25$. Averaging all measurements, we obtain $\left < \log D_{\rm Fe} \right > = -0.74\pm0.25$. This result is also broadly consistent with what was derived in the case of N49;  $\left < \log D_{\rm Fe} \right > = -0.956\pm0.15$ \citep{Dopita16}. 

Finally, the analysis for [\ion{Fe}{7}]$\lambda 6087$/H$\alpha$ is shown in Figure \ref{fig10}. Note that this line requires shock velocities of $v_s > 180$\,km/s in order to be produced with appreciable strength in the radiative shock. Here, the inferred depletion factors range  from -0.0 to $\sim -0.75$ with  $\left < \log D_{\rm Fe} \right > = -0.4$.

Given the intrinsic uncertainties in the atomic data for the forbidden lines of Fe, we can conclude that the [\ion{Fe}{3}] and the [\ion{Fe}{7}] lines imply a mean depletion factor of order $\left < \log D_{\rm Fe} \right > = -0.6\pm0.25$, while the  [\ion{Fe}{2}] lines give a depletion factor $\left < \log D_{\rm Fe} \right > = -0.15\pm0.15$. Thus, in both of these N132D and in N49, we find that Fe-bearing dust grains are not fully destroyed until they reach the recombination zone of the shock. The detail of the grain destruction process were discussed in  \citet{Dopita16}, where it was concluded that this type of depletion pattern supports the grain destruction model of \citet{Seab83} and \citet{Borkowski95}. 

\subsection{[Ni II] diagnostics}
Since we also observe strong [\ion{Ni}{2}]$\lambda 7378$, it is worth asking whether the Ni-containing grains suffer the same fate as the Fe-containing grains. The diagnostic plots for [\ion{Ni}{2}]/H$\alpha$ are given in Figure \ref{fig11}. We find a depletion factor of $\sim -0.4$, intermediate between the [\ion{Fe}{3}] and [\ion{Fe}{3}] results, but again consistent with a large fraction of dust grains being destroyed in the shock.

\subsection{The highly ionised species of Fe}
To produce the highly ionised species of Fe; [\ion{Fe}{9}]${\lambda 8235}$, [\ion{Fe}{10}]${\lambda 6374}$ and [\ion{Fe}{14}]${\lambda 5303}$ requires shock velocities $v_s \gtrsim 200$\,km/s. Indeed, the self-consistent radiative shock models do not produce strong [\ion{Fe}{10}]${\lambda 6374}$ until $v_s \sim 250$\,km/s and [\ion{Fe}{14}]${\lambda 5303}$ becomes bright only for $v_s > 350$\,km/s. The diagnostics of expansion velocity, velocity dispersion, and the spectral diagnostics presented above all point to shock velocities for typical clouds in the range $160$\,km/s$ < v_s < 300$\,km/s. Thus it is clear that the relatively strong [\ion{Fe}{10}] and [\ion{Fe}{14}] lines observed must arise in another region with higher shock velocity. 

This is not unexpected, given the appearance of the clouds on the HST images, which often show a bright core in H$\alpha$, surrounded by fainter filamentary structures. We can identify these regions with the bright radiative shocks being driven into the denser cores of these clouds, and a faster, but only partially-radiative shock driven by a similar external ram-pressure sweeping through the less dense outer regions of the same clouds. Indeed in the case of another SNR in the LMC; N49, both the spatial and dynamical distinction between the [\ion{Fe}{14}]${\lambda 5303}$ and the [\ion{Fe}{2}]${\lambda 5159}$ emission are very marked \citep[cf. Fig.3 in][]{Dopita16}.

The detailed diagnostic diagrams for these two Fe species are shown in Figure \ref{fig12}. For the standard grid, both fail to reproduce the observed strength of the lines, but the [\ion{Fe}{14}] is particularly bad. Introducing a partially-radiative shock leads to a dramatic improvement. The derivation of the plausible shock parameters { for these shocks} is as follows. 

First, we note that the mean cloud shock ram pressure derived from the observations (see Section \ref{cloudshocks}, below) is $P_{\mathrm{ram}} = 3.3\times10^{-7}$\,dynes/cm$^2$. At the mean pre-shock particle density implied for these shocks, $n_0 \sim 240$\,cm$^{-3}$, the mean cooling age to $T_e = 1000$\,K is 400\,yr. This provides an initial estimate of the shock age. Another way is to take the age of the SNR as estimated from the  [\ion{O}{3}] dynamics of the fast-moving material and estimate from the position of the cloud within the SNR how long it has been since the blast wave overran the cloud. In Section \ref{intro} we saw that estimates of the age of N132D vary from 1350 -- 3440\,yr for the SNR, with a probable age of $\sim 2500$\,yr  \citep{Danziger76, Lasker80, Vogt11}. The detailed position of the clouds within the SNR blast wave cannot be reliably estimated from their projected distances from the boundary of the shell, but we will take this as being $\sim 25$\%, as an upper limit, giving a mean shock age of $\lesssim 600$\,yr.

With the shock ram pressure given above, the mean pre-shock density at a given shock velocity is fully determined; for shock velocities of 300, 350 and 400\,km/s the pre-shock particle densities are 160, 117 and 90\,cm$^{-3}$, respectively. None of these shocks can become fully radiative within this timescale. In the 600\,yr available, the 300\,km/s shock cools from $T_e=1.27\times 10^6$\,K to  $T_e=6.06\times 10^5$\,K; the 350\,km/s shock from $T_e=1.72\times 10^6$\,K to  $T_e=1.43\times 10^6$\,K and the 400\,km/s shock from $T_e=2.25\times 10^6$\,K to  $T_e=2.07\times 10^6$\,K.

The optical spectra of these shocks only contain emission lines of hydrogen and helium, as well as lines of coronal species of S, Si, Ar, Fe and Ni. The 300\,km/s shock is too slow to reproduce the observed [\ion{Fe}{14}]/[\ion{Fe}{10}] ratio, while the 400\,km/s overproduces  [\ion{Fe}{14}] relative to [\ion{Fe}{10}], and is somewhat too faint relative to the cloud shock and its precursor. The curves on Figure \ref{fig12} are presented for the 350\,km/s shock. In this figure, we added the contribution of the fast shock to that of the cloud shock, assuming that the fast shock covers an area of either 3 times, or 10 times that of the cloud shock. We assume that dust has been fully destroyed by shattering and sputtering in these fast shocks. Adding this fast shock contribution, we now achieve a satisfactory fit with the observational diagnostics for both [\ion{Fe}{10}]  and  [\ion{Fe}{14}].

The WiFeS images of N132D support the hypothesis that the faster [\ion{Fe}{10}]  and  [\ion{Fe}{14}] - emitting shocks cover a greater area than the cloud shocks which are bright in H$\alpha$. This is shown in Figure \ref{fig13}, in which we have constructed an image in H$\alpha$ (red),  [\ion{Fe}{10}]  (green) and  [\ion{Fe}{14}] (blue). In this image the clouds show up as red or mauve blobs, while the  [\ion{Fe}{14}] emission forms diffuse halos about these clouds, and is much more extensive. In general, the [\ion{Fe}{14}] emission is much more closely spatially correlated with the \emph{Chandra X-Ray Observatory} images \citep{Borkowski07}, an effect which is also seen in both N49 in the LMC \citep{Dopita16} and in the SMC SNR; 1E 0102.2-7219 \citep{Vogt17}.
  
\begin{figure}
  \includegraphics[width = \columnwidth]{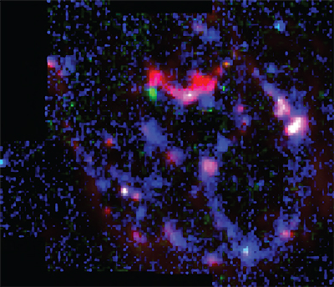}
  \caption{A WiFeS composite image in H$\alpha$ (red),  [\ion{Fe}{10}]  (green) and  [\ion{Fe}{14}] (blue). Note the greater spatial extent of the  [\ion{Fe}{14}] emission, and it close spatial correlation with the X-ray images of N132D. The green blob in this image is not real, but caused by leakage of high-velocity  [\ion{O}{1}]  emission into the [\ion{Fe}{10}] bandpass. } \label{fig13}
 \end{figure}

\section{Detailed Cloud Shock Modelling} \label{cloudshocks}
\subsection{Cloud Shock Parameters}\label{cloudparms}
From the generalised shock diagnostic diagrams, we now turn to the derivation of the shock parameters of the individual shocked ISM clouds in N132D. First, using the shock velocities estimated from both the  [\ion{O}{3}] and He II/He I diagnostic diagrams, the measured  [\ion{S}{2}] electron densities given by the [\ion{S}{2}] $\lambda\lambda 6731/6717$ ratios, and the mean [\ion{S}{2}] electron densities given from the model grid at the same shock velocity, we can estimate the pre-shock particle density, $n_{\mathrm{0}}$, and the ram pressure driving the shock; $P_{\mathrm{ram}}$. These are listed in Table \ref{table5}. For the specific regions P08 and P09, the errors in the estimated shock velocities are great, as are the errors in $n_{\mathrm{0}}$. The region P20 is an incompletely radiative shock so therefore the measured  [\ion{S}{2}] electron density represents an under-estimate of the true density in the  [\ion{S}{2}] recombination zone.

In the final column of Table \ref{table5} we give the time required for the post-shock plasma to cool to 1000\,K,  $ t_{\mathrm{1000K}}$ measured in years. At this point, the optical forbidden lines are no longer emitted in the plasma, and hydrogen is less than 2\% ionised, so the shock is essentially fully radiative. We may conclude that these cloud shocks are somewhat older than $ \left<t_{\mathrm{1000K}}\right> =410$\,yr (excluding P20).

{ The mean ram pressure for the clouds,  $P_{\mathrm{ram}} = 3.1\times10^{-7}$\,dynes/cm$^2$ is much higher than the mean pressure in the X-ray  plasma. This may be estimated using the proton densities and electron temperatures listed in \citet{Williams06}. This gives $7.6 \times 10^{-8}$ dynes cm$^{-2}$ in the NW region, and $5.3 \times 10^{-8}$ dynes cm$^{-2}$ in the S, for an average of $6.4 \times
10^{-8}$ dynes cm$^{-2}$. Thus the ram pressure in the cloud shocks is typically $\sim5$ times higher than in the surrounding X-ray plasma. The reason for this difference is discussed in Section \ref{conc}, below. }

The mean pre-shock density inferred for these clouds is surprisingly high; $\left<n_{\mathrm{0}} \right> = 243$\,cm$^{-3}$ (excluding P08, P09 and P20). However, the pressure in the ISM of the LMC has been estimated using the excited C\,I emission along a number of sight lines by \citet{Welty16}. This analysis gave $\left<\log P/k \right> =4.02$ cm$^{-3}$K. If the N132D clouds were characterised by the same ISM pressure before they were engulfed by the supernova shock wave (which is not at all a certain assumption), then their equilibrium temperature would have been $\sim 40$\,K, and they can be therefore considered to be part of the cold neutral medium phase of the ISM.

 \begin{table}
 \centering
 \small
   \caption{The physical and shock parameters of the interstellar clouds in N132D}
    \label{table5}
   \scalebox{0.9}{
  \begin{tabular}{lccclll}
 \hline
 \hline
Posn.  & $v_s$ & [S II] & $n_{\mathrm{[SII]}}$ &  $n_{\mathrm{0}}$ & $P_{\mathrm{ram}}$ & $ t_{\mathrm{1000K}}$ \\
  \# & km/s & 6731/6717 & cm$^{-3}$ &  cm$^{-3}$ & dynes/cm$^2$ & yr. \\
 \hline
&    &  &  &  & & \\
P01 & $225\pm 40$  & 1.76 & 3760 & 304 & $3.5\times10^{-7}$ & 260  \\
P02 & $208\pm 40$  & 1.83 & 4590 & 365 & $3.6\times10^{-7}$ & 165 \\
P03 & $225\pm 50$  & 1.55 & 2180 & 176 & $2.0\times10^{-7}$ & 455 \\
P04 & $200\pm  50$ & 1.87 & 5180 & 445 & $4.1\times10^{-7}$ & 222 \\
P05 & $263\pm 35$  & 1.66 & 2880 & 198 & $3.2\times10^{-7}$ & 450 \\
P06 &  $210\pm 50$ & 1.76 & 3760 & 293 & $3.3\times10^{-7}$ & 185 \\
P07 & $238\pm 70$  & 1.78 & 3840 & 323 & $4.0\times10^{-7}$ & 240 \\
P08 & $\sim240$       & 1.70 & 3200 & $\sim265$ & $\sim 3.4\times10^{-7}$ & $\sim 370$ \\
P09 & $\sim240$       & 1.55 & 2180 & $\sim180$ & $\sim 2.3\times10^{-7}$ & $\sim 550$ \\
P10 & $245\pm 40$  & 1.68 & 3030 & 222 & $3.4\times10^{-7}$ & 395 \\
P11 & $285\pm  35$ & 1.97 & 7250 & 426 & $5.3\times10^{-7}$ & 365 \\
P12 & $235\pm  25$ & 1.74 & 3560 & 308 & $3.6\times10^{-7}$ & 290 \\
P13 & $250\pm  40$ & 1.45 & 1690 & 129 & $1.8\times10^{-7}$ & 750 \\
P14 & $228\pm 30$ & 1.53  & 2070 & 170 & $2.0\times10^{-7}$ & 450 \\
P15 & $205\pm 25$ & 1.43  & 1610 & 132 & $1.4\times10^{-7}$ & 430 \\
P16 & $238\pm 50$ & 1.78  & 3970 & 288 & $3.9\times10^{-7}$ & 275 \\
P17 & $210\pm 40$ & 1.45  & 1690 & 132 & $1.5\times10^{-7}$ & 410 \\
P18 & $360\pm 20$ & 1.52  & 2020 & 117 & $3.7\times10^{-7}$ & 1100\\
P19 & $310\pm 40$ & 1.53  & 2070 & 131 & $3.1\times10^{-7}$ & 820 \\
P20 & $190\pm 20$ & 1.16  &   760 & 72 & $>5.4\times10^{-8}$ & ... \\

 &    &  &  &  & & \\
 \hline
Av: & 246  &  &  & 243 & $3.1\times10^{-7}$& 410\\
\hline
 \end{tabular}}
\end{table}

\subsection{Chemical Abundances \& Depletion Factors}
We selected the four brightest clouds; P12, P14, P15 and P18 for detailed study, with the aim of deriving more precise shock parameters, and to derive the chemical abundances of the elements in the ISM of the LMC through high-precision shock modelling.

In order to measure the goodness of fit of any particular model, we measure the degree to which it reproduces the density-sensitive [S\,II] $\lambda\lambda 6731/6717$ ratio, and we also seek to minimise the L1-norm for the fit. That is to say that we measure the modulus of the mean logarithmic difference in flux (relative to H$\beta$) between the model and the observations \emph{viz.};
\begin{equation}
{\rm L1} =\frac{1}{m}{\displaystyle\sum_{n=1}^{m}} \left | \log \left[ \frac{F_n({\rm model})} {F_n({\rm obs.)}} \right]  \right |. \label{L1}
\end{equation}

This procedure weights fainter lines equally with stronger lines, and is therefore more sensitive to the values of the input parameters. We simultaneously fit to those emission lines which are most sensitive to the controlling parameters of shock velocity and ram pressure; effectively He\,I, He\,II, [O\,I], [O\,II] and  [O\,III] as well as the [N\,I], [N\,II], and [S\,II] lines. In addition, we estimate the gas-phase heavy element abundances using the bright [Fe\,II] and [Fe\,III] lines, [Fe\,V] 4227\AA\,  [Fe\,VII] 6087\AA\,  [Ni\,II] 7378\AA\, Mg\,I] 4561\AA, Ca\,II 3933\AA\ and [Ca\,II] 7291\AA\ lines. We assume that the more excited species of Fe are depleted in the gas phase by the same amount as Fe\,III. For the reason given above we do not attempt to fit either the  [Fe\,X] 6374\AA\ or the  [Fe\,XIV] 5303 \AA\ lines, given that they arise in a separate phase of the ISM. Other elements such as Ne, Ar and Cl are fit using the [Ne\,III] 3969\AA\, [Ar\,III] 7136\AA\ and [Cl\,II] 8579\AA\ lines. 

Our models simultaneously { optimise the fit to} 40 emission lines, and we manually iterate shock velocity, ram pressure and chemical abundances until the fit is optimised. The results for these four clouds are shown in Table \ref{table6}. The L1-norm provides a good estimate for the error on the abundance of those elements fitted with only one or two emission lines. For O, N and S the accuracy is higher. In Table 2, we have adopted the abundances of Mg, Ca, Fe and Ni from \citet{Russell92} in order to estimate the depletion factors of these elements.

The best fit was achieved without the inclusion of the precursor emission, which in principle could account for as much as 25\% of the total emission. There are two possible reasons for this. First, the finite extent of the cloud may limit the length of the precursor, which is maximised in our plane-parallel models. Second, the local curvature of the shock front may allow significant escape of the upstream ionising UV photons. This is suggested by the fact that our models of these four clouds systematically overestimate the \ion{He}{2}$\lambda 4686$/ \ion{He}{1}$\lambda 5876$ ratio, which suggests that the incoming plasma is over-ionised by the computed precursor radiation field given by the model.

 \begin{table}
 \centering
 \small
   \caption{Goodness of fit, shock parameters, ionic depletions and chemical abundances, $12+\log(\mathrm{[X/H]})$,  for  clouds in N132D.}
    \label{table6}
   \scalebox{0.9}{
  \begin{tabular}{lrrrr}
\hline
\hline
Posn: & P12 & P14 & P15 &  P18  \\
  \hline
&    &  &  &  \\
L1-norm &  0.059  & 0.083 &  0.086 &  0.081 \\
$v_s$ (km/s) &  $210\pm 25$ & $225\pm  50$ &  $210\pm30$ &   $260\pm30$ \\
$n(\mathrm H)_0$ (cm$^{-3}$) &$410\pm30$  & $325\pm25$ & $160\pm25$ & $216\pm25$  \\
$P_{ram}$ &  3.88 & 3.57 &  1.51 &  2.90  \\
($\times 10^7$dynes cm$^{2}$)&    &  &  &  \\
&    &  &  &  \\
  \hline
$\log D_{\mathrm{FeII}}$ & -0.13 & -0.13 & -0.13 & -0.13 \\
$\log D_{\mathrm{FeIII}}$ & -0.67 & -0.63 & -0.58 & -0.47 \\
$\log D_{\mathrm{MgII}}$ & -0.94 &  -0.57 & -1.33 & -0.74 \\
$\log D_{\mathrm{CaII}}$ & -0.30 & -0.05  & -0.06 & -0.04 \\
$\log D_{\mathrm{NiII}}$ &  -0.69 &  -0.28 &-0.30 & -0.40 \\
  \hline
&    &  &  &  \\
He & 10.84 & 10.86 & 10.85 &  10.86 \\
N & 7.11  & 7.08 &  7.15 &  7.15  \\
O &  8.28 &  8.30 & 8.32 &   8.32 \\
Ne &  7.45 & 7.42 & 7.45 & 7.45 \\
Mg &  7.47 & 7.47 &  7.47 & 7.47 \\
S &  6.95  & 7.10 &  7.00 & 7.00 \\
Cl &  4.35  & 4.90 &  4.35 & 5.03 \\
 Ar &  6.13  & 5.65  &  5.68 & 5.68 \\
 Ca &  5.97  & 5.97  & 5.97 & 5.97\\
 Fe &  7.23  & 7.23  & 7.23 & 7.23 \\
 Ni &   5.96 & 5.96  & 5.96 & 5.96 \\
\hline
\hline
 \end{tabular}}
\end{table}

{ The derived depletion factors for Mg given in Table \ref{table6}  are likely to have very large, and unquantifiable errors attached to them. There is only one Mg line detected, and it is produced by neutral Mg in the tail of the recombination zone of the shock. Given the low ionisation potential of this species (7.64\,eV), the Mg\,I] line strength will be highly dependent on the unknown stellar UV radiation field, which by ionising this species would depress the line strength and lead to a larger depletion factor being inferred.}

For the remaining elements we derive the following mean LMC interstellar abundances, $12+\log(\mathrm{[X/H]})$,  of He: $ 10.86\pm 0.05$, N: $ 7.12\pm 0.07$, O: $ 8.31\pm 0.04$, Ne: $ 7.44 \pm 0.08$, S: $ 7.01\pm 0.06$, Cl: $ 4.66\pm 0.11$ and Ar: $ 5.78\pm 0.11$. These abundances are generally similar to those obtained in our detailed shock fitting of the supernova remnant N49 \citep{Dopita16}; He: 10.92, N: 7.10, O: 8.46, Ne: 7.79, S: 7.05, Cl: 5.2 and Ar: 6.1.

These abundances should also be compared with those from  \citet{Russell92}: He: $ 10.94\pm 0.03$, N: $ 7.14\pm 0.15$, O: $ 8.35\pm 0.06$, Ne: $ 7.61\pm 0.05$, S: $ 6.70\pm 0.09$, Cl: $ 4.76\pm 0.08$ and Ar: $ 6.29\pm 0.25$. Given the differences in the methodology used to derive these, and the changes in key atomic input parameters over the years, the agreement between these two is as good as could be expected. No systematic difference is seen for the elements most important in the modelling; N and O, but the derived He, Ne, S, Cl and Ar abundances are only marginally consistent with each other. What is now required is a re-derivation of the abundances using H\,II regions and the same modelling code to compare with the SNR-derived abundances.

The derived abundances can also be compared to more recent determinations from stellar observations. \citet{Korn05} gave an new analysis of a slow-rotating B star in NGC\,2002 which gave N: $ 6.99\pm 0.2$, O: $ 8.29\pm 0.2$, and Mg: $ 7.44\pm 0.2$. These values differed very little from his earlier work \citep{Korn02}.

With respect to the refractory elements, \citet{Korn00} derived appreciably lower Mg and Fe abundances than \citet{Russell92}; Mg: $ 6.96\pm 0.22$ and Fe: $ 7.09\pm 0.15$. Using these values, our results are consistent with full destruction of the Fe-bearing grains in the [\ion{Fe}{2}]--emitting zone, and the estimated Mg depletion factor is lowered to only $\log D_{\mathrm{MgII}} = -0.38$. However, the later \citet{Korn02} paper gave Mg: $7.37\pm 0.06$ and Fe: $ 7.33\pm 0.0.03$, which is in much closer agreement with the \citet{Russell92} estimates.

In conclusion, we can be fairly confident of the LMC He, N and O abundances; He: $ 10.92\pm 0.04$, N: $ 7.12\pm 0.09$ and O: $ 8.32\pm 0.06$. For the other non-refractory elements the recommended values are Ne: $ 7.52\pm 0.09$, S: $ 7.00\pm 0.15$, Cl: $ 4.70\pm 0.08$ and Ar: $ 5.78\pm 0.25$.

 Our results are consistent with nearly full destruction of the Fe--containing grains in these 210--260\,km/s shocks, but the Mg--containing grains appear to be only partially destroyed. This argues for a different carrier for the two metals, possibly iron(II) oxide (FeO) for iron and  the magnesium silicates, forsterite (Mg$_2$SiO$_4$) and enstatite (MgSiO$_3$) in the case of Mg. 
 
 The destruction of  Fe--containing grains appears to have proceeded to a greater extent in N132D than in N49 \citep{Dopita16}. Although the estimated shock velocities are similar (210--260\,km/s in N132D and 200--250\,km/s in N49), the higher ram pressures and higher pre-shock densities in N132D (160--410\,cm$^{-3}$) compared with N49 ($\sim 80$\,cm$^{-3}$) has facilitated this greater fractional dust destruction.
 
 In both N49 and in N132D, the grain destruction in the faster, partially-radiative shocks producing [\ion{Fe}{10}] and [\ion{Fe}{14}] emission is much more advanced. In both these SNR, the shock velocity in these highly-ionised zones is $\sim 350$\,km/s. Thus, thermal sputtering and non-thermal sputtering due to the high relative velocity of the dust grains relative to the hot-post shock plasma \citep{Barlow78, McKee87, Seab87,Jones94} dominates the grain destruction in these fast shocks, while gyro-acceleration (aka betatron acceleration) caused by the compression of the magnetic field in the cooling zone of the shock resulting in grain shattering and vaporisation \citep{Spitzer76, Borkowski95} appears to dominate as the grain destruction process in the slower, denser shocks. 
 
\begin{figure}
  \includegraphics[width = \columnwidth]{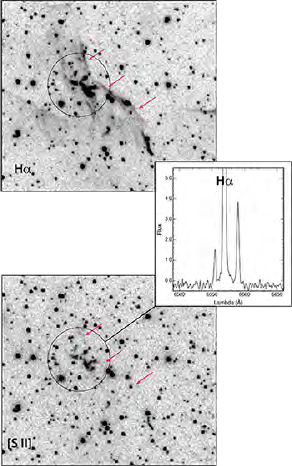}
  \caption{The region P08 as seen by HST in H$\alpha$ (top panel) and in [\ion{S}{2}] (lower panel). The WiFeS extraction aperture is shown as a black circle, and the resultant spectrum around H$\alpha$ is shown in the inset. The Flux units are the same as in Figure \ref{fig4}. Note the broad wings of H$\alpha$ extending out to $\pm900$\,km/s. This broad emission is presumably originating from the filaments which are marked with red arrows on the images. } \label{fig14}
 \end{figure}

\subsection{The partially-radiative region P20}
We already noted in Section \ref{diagnostics} that region P20 has anomalously strong [\ion{O}{3}] emission -- see Figure \ref{fig7}. Furthermore, the [\ion{S}{2}] lines indicate a low density compared with the other regions which produces the low value of the ram pressure given in Table \ref{table5}. All of these point to a shock which has not yet become fully radiative. In this case, it appears that the photoionised precursor is brighter in the optical lines than the shock itself. The extremely strong [\ion{O}{3}] would arise in this precursor.

The observed spectrum of this region proved very difficult to model. Our ``best" model has a relatively poor fit, with an L1-norm of 0.21. This model has a $\sim 400$\,yr old shock with  $v_s = 300$\,km/s moving into a medium with a pre-shock H density $n_0 = 235$\,cm$^{-3}$. In this shock the plasma cools from an initial temperature of $T_e =1.2\times10^6$\,K to a final temperature of  $T_e =8.2\times10^5$\,K. The precursor material is assumed to have no dust destruction with $\log D_{Fe} = -1.00$, and the density of the pre-shock gas $n_0 = 235$\,cm$^{-3}$, and is illuminated by the UV radiation field of the partially-radiative shock in plane-parallel one-sided geometry. This gives a precursor thickness of $\sim 0.12$\,pc (at the point where H is 50\% ionised). For this configuration, the shock produces only 4.5\% of the total H$\beta$ luminosity. Table \ref{table7} provides  comparison of the key emission lines predicted by the model, compared with the observations. 
 \begin{table}
 \centering
 \small
   \caption{The partially-radiative shock model fit to the observations of P20 (H$\beta =100$).}
    \label{table7}
   \scalebox{0.9}{
  \begin{tabular}{lccc}
\hline
\hline
 $\lambda$ (\AA) & ID & Obs. & Model \\
 & & & \\
\hline
3727,9 & [O II] & 419 &  555 \\
3867 & [Ne III] & 82 &  62 \\
3835 & H$\eta$ & 6 &  7 \\
3967 & [Ne III] & 25 &  19 \\
3970 & H$\epsilon$ & 15 &  15 \\
4969,76 & [S II] & 9 &  8 \\
4340 & H$\gamma$ & 45 &  46\\
4363 & [O III] & 84 & 7 \\
4696 & He II & 14 &  35\\
4861 & H$\beta$ & 100 &  100\\
4959 & [O III] & 326 & 165 \\
5007 & [O III] & 967 & 476 \\
5755 & [N II] & 8 & 2 \\
5876 & He I & 6 & 6 \\
6548 & [N II] & 33 & 21 \\
6563 & H$\alpha$ & 294 &  318 \\
6584 & [N II] & 95 & 63 \\
6717 & [S II] & 121 & 83 \\
6731 & [S II] & 142 & 62 \\
7136 & [Ar III] & 17 & 8 \\
7319 & [O II] & 46 & 10 \\
7329 & [O II] & 25 & 8\\
\hline
\hline
 \end{tabular}}
\end{table}

Note the extreme discrepancy in the predicted strength of the [\ion{O}{3}] $\lambda 4363$ line. From the model, the predicted temperature in the \ion{O}{3} zone is $T_e = 13600$\,K, while the data suggest an electron temperature of $T_e \sim 49000$\,K in this zone. Similar, but smaller, temperature discrepancies are suggested for the  \ion{O}{2} and  \ion{N}{2} zones. This requires another form of heating in the precursor such as electron conduction,  cosmic ray heating or else ionisation by the general EUV radiation field of the nebula.  Apart from these temperature discrepancies, the [\ion{S}{2}] $\lambda\lambda 6731/6717$ ratio also requires a higher electron density; $n_{\mathrm{[SII]}} \sim 760$\,cm$^{-3}$. \newpage

\subsection{The Balmer-Dominated Shocks in P08}
We noted above that the region P08 also displays an anomalous spectrum, deficient in the forbidden lines with respect to the Balmer lines, with the [\ion{O}{3}] lines being particularly weak -- see Figure \ref{fig7}. The cause of this appears to be ``non-radiative'' or Balmer-dominated shocks. These occur when a fast shock runs into a cold, essentially un-ionised ISM. A narrow component of the Balmer lines arises from direct collisional excitation of Hydrogen by the fast electrons, and a broad Balmer component is produced by charge exchange with fast protons behind the shock and subsequent collisional excitation of the fast neutral hydrogen resulting by the hot electrons \citep{Chevalier78, Chevalier80}  -- see the reviews by \citet{Heng10} and \citet{ Ghavamian13}, and references therein.
In Figure \ref{fig14}, we show the HST view of the P08 region, and the spectrum extracted from the WiFeS data cube. It is most likely that the broad component of H$\alpha$ seen in the spectroscopy arises from the filaments which are visible in the H$\alpha$ image, but not in the [\ion{S}{2}] image. A broad component is also weakly detected in H$\beta$. This would add N132D to the select group of SNR in which this has been detected -- which includes Tycho, SN\,1006, RCW\,86, the Cygnus Loop and N103B and SNR 0519-69.0 (DEM N71) in the LMC \citep{Tuohy82, Ghavamian01,  Ghavamian02, Ghavamian17}. { Most, but not all, of these remnants are from Type Ia supernova explosions, since these do not produce a burst of strong EUV radiation to pre-ionise the pre-shock medium. However, if the ISM is shielded from the EUV flash in some way, or if it is dense enough to recombine in the period between the explosion and the arrival of the shock, then nothing precludes the formation of a Balmer-dominated shock. Given the inferred age of the SNR, the recombination condition would imply  that the shock is passing through a  medium with a density $ \gtrsim 20$\,cm$^{-3}$. The width of the H$\alpha$ line suggests that the shock velocity in the Balmer-dominated shock is of order $\sim900$\,km/s. This is similar to the blast wave velocity derived from the X-ray data \citep{Favata97, Hughes98, Borkowski07}, so it seems extremely likely that the broad Balmer lines arise in the blast wave itself. }

\section{Conclusions} \label{conc}
The environment of N132D provides an excellent sampling of shocked clouds in the Bar of the LMC. The typical physical size of these clouds is $\sim 1.0$\,pc. Given that they have a typical pre-shock density of $\sim240$\,cm$^{-3}$ (from Table \ref{table5}), we can infer cloud masses of few solar masses. Using the images from Figure \ref{fig3} and the pre-shock densities from Table \ref{table5}, we obtain masses in the range $0.1- 20$\,M$_{\odot}$ with a mean of $\sim 4$\,M$_{\odot}$. Thus we infer these clouds initially represented  typical ISM self-gravitating Bonnor-Ebert spheres such as those recently investigated on a theoretical basis by  \citet{Sipila11}, \citet{Fischera14} and \citet{Sipila17}. Given that the shock which moves into them following the passage of the supernova blast-wave is strongly compressive, and that such Bonnor-Ebert spheres can be marginally stable against collapse, it is tempting to imagine that the supernova shock may later induce formation of $\sim 1.0 M_{\odot}$ stars within the cores of those clouds.  

{ Now, let us consider the discrepancy between the thermal pressure measured by the X-rays, and the ram pressure driving the cloud shocks, pointed out in Section \ref{cloudparms}. The X-ray plasma has a thermal pressure of $P_{\mathrm{therm}} = 6.4 \times 10^{-8}$ dynes cm$^{-2}$, while the ram pressure in the clouds is $P_{\mathrm{ram}} = 3.1 \times 10^{-7}$ dynes cm$^{-2}$. However, the pressure driving the cloud shocks is provided by the stagnation pressure behind the cloud bow shock or bow wave produced in the expanding thermal plasma which fills the SNR. This is always greater than the pressure in the pre-shock hot thermal plasma. The geometry of this interaction is pictured in \citet{McKee75, Hester86}} and \citet{Farage10}. In the framework of a plane-parallel strong shock engulfing the cloud the stagnation pressure is about twice as large as the thermal pressure \citep{McKee75, Hester86} and this is only very weakly dependent to the Mach number of the primary blast wave (see \citet{Hester86}, Table 2).

The X-ray data enable us to estimate the stagnation pressure in two ways. First, we may use the Sedov theory to estimate the blast wave velocity. Using the figures given in \citet{Hughes98} for the explosion energy, age and pre-shock density in the SNR, we derive the blast-wave velocity, $v_{\mathrm B} = 830$\,km/s. Alternatively, we can use the measured thermal plasma temperatures to derive the blast wave velocity, using the relation $T_{\mathrm B} =(3 \mu m_H/16k)v_{\mathrm B}^2$, where $m_H$ is the pre-shock hydrogen density and the molecular weight $\mu$ is appropriate to a fully-ionised plasma. Estimates of the thermal temperature are in the range $0.68 < kT < 0.8$\,keV \citep{Favata97, Hughes98, Borkowski07}, which implies a blast wave velocity of $\sim 760$\,km/s. Therefore, the ram pressure associated with expansion is $\sim 4 \times 10^{-8}$ dynes cm$^{-2}$. Adding this to the thermal pressure gives an estimate of the stagnation pressure, $\sim 1.0 \times 10^{-7}$ dynes cm$^{-2}$. 

The remaining difference between the ram pressure of the cloud shocks and the estimated stagnation pressure is a factor of three. This difference may be accounted for by the fact that the cloud shocks are convergent towards the centre of mass, a possibility raised in the context of the Cygnus Loop SNR by \citet{Hester86} as a promising mechanism for increasing pressure in shocked dense clouds. For shocks in self-gravitating isothermal spheres such as those considered here, families of self-similar solutions have been obtained by \citet{Lou14}. In such shocks, the energy density increases as the shock moves toward the center of the cloud, increasing the ram pressure. This effect can easily account for the difference between the measured ram pressure of the cloud shocks and the estimated stagnation pressure. Additionally, convergent cloud shocks are unstable in the presence of small deviations from sphericity \citep{Kimura90}, and elongated clouds may break up into separate globules \citep{Kimura91}. Such instabilities are the likely reason for the complex morphologies of the shocks in individual clouds seen in Figure \ref{fig3}.}

Because the shocked clouds of N132D represent gravitationally-confined samples of the ISM as they existed before the supernova event, and because these clouds are dense enough that they would not be appreciably affected in their chemical composition by any pre-supernova mass-loss, they present ideal samples of pristine ISM in the LMC. Our radiative shock  analysis has enabled us to estimate accurate gas-phase chemical abundances for a number of elements.

From our model grid, we have demonstrated that the [\ion{O}{3}] $\lambda\lambda 4363/5007$ ratio is a good indicator of shock velocity, and that the \ion{He}{2}/\ion{He}{1}  $\lambda\lambda 4686/5876$ ratio may also be used for this purpose, although it is rather less reliable. Typical shock velocities are $\sim240$\,km/s, in agreement with the shock velocities inferred from the kinematics and the measured emission line widths. Using these emission line diagnostics, we have analysed the depletion onto dust in the various ionic stages of Fe, and in \ion{Mg}{2} \ion{Ca}{2} and \ion{Ni}{2}. In common with the SNR N49, we find that dust has been mostly destroyed in the region emitting the [\ion{Fe}{2}] lines, while a smaller fraction has been destroyed in the \ion{Fe}{3} and \ion{Fe}{7} zones, consistent with the grain destruction models of \citet{Seab83} and \citet{Borkowski95}. However, \ion{Mg}{2} shows an appreciably higher depletion factor, suggesting that the Mg silicates are more resistant to destruction than the Fe-bearing grains.

It is clear that the highly ionised species of iron;  \ion{Fe}{10} and  \ion{Fe}{14} originate in faster, partially radiative, filamentary and spatially extensive shocks surrounding the dense clouds. Some of these shocks may well be part of the primary blast wave of the SNR. In these, most of the refractory elements have been destroyed by thermal sputtering.

 \begin{table}
 \centering
 \small
   \caption{Recommended LMC chemical abundances.}
    \label{table8}
   \scalebox{1.0}{
  \begin{tabular}{lc}
\hline
\hline
Element & $12+\log(\mathrm{[X/H]})$ \\
\hline
&   \\
He & $10.92\pm 0.04$ \\
C  & $8.06\pm 0.09$ \\
N  & $7.12\pm 0.09$ \\
O  & $8.32\pm 0.06$ \\
Ne & $7.52\pm 0.09$ \\
Mg & $7.37\pm 0.06$ \\
Si  & $7.10\pm 0.07$ \\
S &  $7.00\pm 0.15$ \\
Cl & $4.70\pm 0.08$ \\
Ar & $5.78\pm 0.25$ \\
Fe  & $7.33\pm 0.03$ \\
\hline
\hline
 \end{tabular}}
\end{table}

A detailed shock analysis of the four brightest clouds has allowed us to determine the chemical abundances of a number of elements. Comparing these with values given earlier, and with stellar abundance determinations we can now provide a ``recommended''  set of LMC abundances, which we present here in Table \ref{table8}. Here, for completeness, the abundances of C, Mg, Si and Fe have been taken from the work of \citet{Korn02} using Magellanic Cloud B-stars in NGC\,2004.

Finally we have identified two anomalous regions, P08 and P20. The former contains a contribution from fast  Balmer-dominated shocks, while the latter represents an unusual partially-radiative shock, dominated by precursor emission which seems to be heated by an unknown source to very high electron temperatures.

\section*{Acknowledgments}
MD and RS acknowledge the support of the Australian Research Council (ARC) through Discovery project DP16010363. Parts of this research were conducted by the Australian Research Council Centre of Excellence for All Sky Astrophysics in 3 Dimensions (ASTRO 3D), through project number CE170100013. FPAV and IRS thank the CAASTRO AI travel grant for generous support. IRS was supported by the ARC through the Future Fellowship grant FT1601000028. AJR is supported by the Australian Research Council through Future Fellowship grant FT170100243.

This research has made use of \textsc{matplotlib} \citep{Hunter07}, \textsc{astropy}, a community-developed core \textsc{python} package for Astronomy \citep{AstropyCollaboration13}, \textsc{APLpy}, an open-source plotting package for \textsc{python}\citep{Robitaille12}, and \textsc{montage}, funded by the National Science Foundation under Grant Number ACI-1440620 and previously funded by the National Aeronautics and Space Administration's Earth Science Technology Office, Computation Technologies Project, under Cooperative Agreement Number NCC5-626 between NASA and the California Institute of Technology. 

This research has also made use of \textsc{drizzlepac}, a product of the Space Telescope Science Institute, which is operated by AURA for NASA, of the \textsc{aladin} interactive sky atlas \citep{Bonnarel00}, of \textsc{saoimage ds9} \citep{Joye03} developed by Smithsonian Astrophysical Observatory, of NASA's Astrophysics Data System, and of the NASA/IPAC Extragalactic Database \cite{Helou91} which is operated by the Jet Propulsion Laboratory, California Institute of Technology, under contract with the National Aeronautics and Space Administration. Support for MAST for non-HST data is provided by the NASA Office of Space Science via grant NNX09AF08G and by other grants and contracts. This research has also made use of NASA's Astrophysics Data System.

This work has made use of data from the European Space Agency (ESA) mission {\it Gaia} (\url{https://www.cosmos.esa.int/gaia}), processed by the {\it Gaia} Data Processing and Analysis Consortium (DPAC, \url{https://www.cosmos.esa.int/web/gaia/dpac/consortium}). Funding for the DPAC has been provided by national institutions, in particular the institutions participating in the {\it Gaia} Multilateral Agreement. Some of the data presented in this paper were obtained from the Mikulski Archive for Space Telescopes (MAST). STScI is operated by the Association of Universities for Research in Astronomy, Inc., under NASA contract NAS5-26555. Support for MAST for non-HST data is provided by the NASA Office of Space Science via grant NNX09AF08G and by other grants and contracts.


\bibliographystyle{aasjournal}

\clearpage
\setcounter{table}{0}
\renewcommand{\thetable}{A\arabic{table}}
 \section*{Appendix}
In the following four Tables,  \ref{tableA1} through \ref{tableA4}, we present the measured line fluxes of the 20 shocked ISM clouds in N132D. No reddening correction has been applied, since the mean Balmer Decrement as measured;  H$\alpha$/H$\beta$/H$\gamma$/H$\delta$ = 298.7/100.0/44.7/23.9 is indistinguishable (within the errors) with the theoretical Balmer Decrement for $\log n_e =10^4$\,cm$^{-3}$, and $T_e = 5000$\,K; 300/100/46.0/25.3.  On the basis of the soft X-ray absorption \citet{Borkowski07} determined a best-fit hydrogen column density of $1.4\times10^{20}$ cm$^{-2}$ in the western part of N132D and $1.4-4.1\times10^{21}$ cm$^{-2}$ in the south. Assuming dust is not destroyed in the bar of the LMC, and using the \citet{Fitzpatrick86} $N_{\mathrm H}/E_{\mathrm{B - V}}$ conversion factor, these figures would imply a reddening of $E_{\mathrm{B - V}} = 0.005$\,mag. and 0.06--0.17\,mag., respectively. 
\begin{table*}
 \centering
 \small
   \caption{Measured line fluxes for N132D shocked cloudlets P01 -- P05}
    \label{tableA1}
   \scalebox{0.5}{
  \begin{tabular}{lccccccccccc}
 \hline
\bf{Cloud ID\#:} &  & \bf{P01} &  & \bf{P02} &  & \bf{P03} &  & \bf{P04} &  & \bf{P05} & \\
Lambda & ID & Flux & Error & Flux & Error & Flux & Error & Flux & Error & Flux & Error\\
 \hline
3736.02 & [O II] & 176.1 & 10.4 & 102 & 7.9 & 247.2 & 23.0 & 119.1 & 2.7 & 191.2 & 10.6\\
3728.82 & [O II] & 92.8 & 9.8 & 79.4 & 8.7 & 278.4 & 23.4 & 46.1 & 1.7 & 59.6 & 6.5\\
3835.4 & H$\eta$ & 13.7 & 7.2 & 18.1 & 7.0 & 6.9 & 3.4 & 11.6 & 2.0 & ... & \\
3868.76 & [Ne III] & 19.1 & 2.3 & 27.0 & 2.3 & 60.8 & 4.2 & 21.7 & 1.4 & 33.5 & 2.8\\
3888.8 & HeI, H & 27.0 & 5.9 & 29.4 & 4.0 & 15.5 & 3.9 & 19.6 & 1.4 & 22.8 & 2.5\\
3933.66 & Ca II & 17.3 & 3.5 & 41.6 & 8.8 & 13.9 & 2.1 & 13.5 & 1.7 & 19.7 & 4.5\\
3969.0 & [Ne III], Ca II, H$\epsilon$ & 42.1 & 4.2 & 44.4 & 5.1 & 41.5 & 5.5 & 29.3 & 1.6 & 31.2 & 3.4\\
4068.6 & [S II] & 34.6 & 1.9 & 27.9 & 1.5 & 33.8 & 2.7 & 31.6 & 0.8 & 35.3 & 2.3\\
4076.35 & [S II] & 14.9 & 1.8 & 11.2 & 1.6 & 9.6 & 4.1 & 8.7 & 0.8 & 13.5 & 2.8\\
4101.74 & H$\delta$ & 28.8 & 2.2 & 27.4 & 2.1 & 22.3 & 2.4 & 21.4 & 1.3 & 32.9 & 3.9\\
4244.0 & [Fe II] & 7.6 & 2.5 & 3.2 & 0.9 & ... &  & 4.8 & 0.5 & ... & \\
4276.83 & [Fe II] & ... &  & 2.3 & 1.3 & ... &  & 2.5 & 0.8 & ... & \\
4287.43 & [Fe II] & 3.7 & 1.0 & 2.1 & 0.9 & 2.8 & 1.6 & 6.4 & 0.6 & ... & \\
4340.47 & H$\gamma$ & 57.5 & 2.3 & 41.3 & 1.5 & 49.4 & 1.7 & 41.9 & 0.9 & 42.9 & 2.2\\
4359.14 & [Fe IX] & 2.9 & 1.3 & ... &  & ... &  & ... &  & ... & \\
4363.21 & [O III] & 9.9 & 2.8 & 8.5 & 1.5 & 14.6 & 3.4 & 8.0 & 1.8 & 7.6 & 1.2\\
4416.27 & [Fe II] & 5.0 & 1.7 & 8.2 & 1.7 & 7.7 & 1.9 & 7.3 & 0.9 & ... & \\
4452.38 & [Fe II] & 2.2 & 0.6 & ... &  & ... &  & 4.5 & 1.2 & ... & \\
4457.95 & [Fe II] & ... &  & ... &  & ... &  & 1.3 & 0.3 & ... & \\
4471.5 & He I & 3.8 & 1.1 & 5.9 & 2.5 & 12.2 & 2.9 & 5.5 & 0.6 & 2.4 & 1.1\\
4658.5 & [O III],[Fe III] & 1.6 & 0.5 & 6.5 & 0.8 & 5.2 & 1.1 & 7.2 & 0.6 & 2.6 & 0.9\\
4685.7 & He II & 7.0 & 1.9 & 9.4 & 1.1 & 3.2 & 1.2 & 6.8 & 0.6 & 5.6 & 1.6\\
4701.53 & [Fe III] & 0.8 & 0.8 & ... &  & ... &  & 3.4 & 0.6 & ... & \\
4754.68 & [Fe III] & 2.0 & 0.6 & 5.5 & 2.3 & ... &  & 1.4 & 0.6 & ... & \\
4814.54 & [Fe II] & 1.4 & 0.9 & 11.2 & 2.6 & ... &  & 4.3 & 0.5 & 2.4 & 1.6\\
4861.29 & H$\beta$ & 100.0 & 1.1 & 100.0 & 1.0 & 100.0 & 2.0 & 100.0 & 0.8 & 100.0 & 1.9\\
4881.0 & [Fe III] & 2.5 & 0.9 & 10.3 & 3.9 & ... &  & 2.2 & 0.4 & ... & \\
4889.62 & [Fe II] & 1.2 & 0.6 & 6.8 & 2.1 & 3.5 & 1.4 & 2.8 & 0.5 & ... & \\
4905.35 & [Fe II] & 1.2 & 0.9 & 4.8 & 1.1 & ... &  & ... &  & ... & \\
4958.91 & [O III] & 20.7 & 1.3 & 27.0 & 1.3 & 74.3 & 2.3 & 31.7 & 0.6 & 49.1 & 1.6\\
5006.84 & [O III] & 52.6 & 1.5 & 97.9 & 1.4 & 198.4 & 4.1 & 95.1 & 1.0 & 157.1 & 1.7\\
5111.64 & [Fe II] & 0.1 & 0.5 & ... &  & ... &  & 4.1 & 1.2 & ... & \\
5158.79 & [Fe II], [Fe VII] & 4.8 & 0.9 & 11.2 & 0.9 & 10.3 & 1.4 & 18.5 & 0.9 & 11.5 & 1.8\\
5198.8 & [N I] & 0.6 & 0.5 & 5.6 & 1.7 & ... &  & 11.5 & 4.7 & ... & \\
5261.66 & [Fe II] & 1.4 & 0.8 & 5.0 & 1.5 & 1.4 & 0.7 & 6.9 & 0.6 & 4.2 & 1.3\\
5270.4 & [Fe III] & 2.0 & 1.1 & ... &  & ... &  & 8.6 & 2.4 & 6.4 & 2.2\\
5273.36 & [Fe II] & 2.1 & 2.0 & 39.3 & 17.8 & ... &  & ... &  & ... & \\
5303.06 & [Fe XIV] & 2.4 & 1.2 & 12.0 & 1.3 & 2.4 & 0.8 & 5.5 & 0.6 & 1.5 & 0.6\\
5334.5 & [Fe II],[Fe VI] & 1.2 & 0.6 & 6.6 & 2.4 & ... &  & 4.5 & 0.5 & 4.2 & 1.4\\
5376.0 & [Fe II],[Fe VIII] & 3.5 & 1.5 & ... &  & ... &  & 2.7 & 0.4 & ... & \\
5527.3 & [Fe II] & ... &  & ... &  & ... &  & 2.4 & 0.7 & 3.1 & 1.4\\
5754.59 & [N II] & ... &  & ... &  & 0.7 & 0.3 & 1.1 & 0.2 & ... & \\
5875.61 & He I & 7.4 & 1.1 & 9.3 & 1.2 & 8.0 & 0.9 & 11.2 & 0.4 & 7.4 & 0.9\\
6086.97 & [Fe VII] & ... &  & ... &  & ... &  & 2.2 & 0.4 & 2.2 & 1.0\\
6300.3 & [O I] & 87.7 & 3.5 & 106 & 4.3 & 62.3 & 1.9 & 90.8 & 1.5 & 73.6 & 3.6\\
6312.06 & [S III] & ... &  & 2.0 & 1.0 & ... &  & 0.8 & 0.4 & 0.5 & 0.2\\
6363.78 & [O I] & 29.4 & 1.4 & 17.8 & 1.3 & 26 & 1.5 & 30.7 & 0.4 & 22.8 & 2.4\\
6374.52 & [Fe X] & 2.3 & 0.5 & 6.5 & 1.3 & 3.9 & 1.2 & 4.2 & 0.4 & ... & \\
6548.05 & [N II] & 28.2 & 1.3 & 19.6 & 0.8 & 32.6 & 2.3 & 29.9 & 0.5 & 28.3 & 1.1\\
6562.82 & H$\alpha$ & 302.3 & 1.4 & 301.2 & 1.9 & 300.9 & 3.1 & 299.5 & 0.9 & 299.5 & 1.6\\
6583.45 & [N II] & 85.0 & 0.9 & 67.0 & 0.8 & 98.9 & 1.5 & 94.8 & 0.4 & 87.0 & 1.2\\
6678.15 & He I & 2.6 & 0.4 & 3.8 & 1.2 & 2.6 & 0.4 & 4.0 & 0.2 & 1.7 & 0.9\\
6716.44 & [S II] & 70.6 & 0.9 & 43 & 0.8 & 83.1 & 2.0 & 60.3 & 0.4 & 67.6 & 0.9\\
6730.82 & [S II] & 124.1 & 1.0 & 78.9 & 0.9 & 128.7 & 2.4 & 113.0 & 0.3 & 112.0 & 1.0\\
7065.25 & He I & 4.4 & 0.5 & 4.4 & 1.1 & 3.9 & 0.5 & 5.3 & 0.2 & 5.1 & 0.7\\
7135.79 & [Ar III] & 1.4 & 0.2 & 4.9 & 0.7 & 6.0 & 0.5 & 4.5 & 0.2 & 4.9 & 0.4\\
7155.1 & [Fe II] & 15.3 & 0.4 & 16.1 & 0.6 & 13.4 & 0.6 & 26.7 & 0.3 & 13.1 & 0.4\\
7172 & [Fe II] & 1.8 & 0.3 & 5.3 & 0.8 & 5.7 & 3.0 & 4.2 & 0.3 & 2.9 & 0.3\\
7291.47 & [Ca II] & 24.3 & 0.5 & 21.6 & 0.9 & 23.7 & 1.0 & 38.6 & 0.4 & 17.9 & 1.0 \\
7321.5 & [Ca II], [O II] & 40.0 & 1.2 & 38.0 & 2.6 & 38.5 & 2.4 & 51.8 & 1.6 & 37.8 & 1.8\\
7330 & [O II] & 11.0 & 0.7 & 17.0 & 1.3 & 18.6 & 1.7 & 18.1 & 1.1 & 23.6 & 1.8\\
7377.82 & [Ni II] & 8.6 & 0.5 & 8.9 & 1.1 & 7.1 & 0.9 & 11.9 & 0.3 & 4.1 & 0.6\\
7388.2 & [Fe II] & 2.3 & 0.4 & 4.6 & 1.3 & ... &  & 3.4 & 0.3 & ... & \\
7411.61 & [Ni II] & 1.1 & 0.4 & 1.7 & 0.8 & ... &  & 1.0 & 0.4 & ... & \\
7452.5 & [Fe II] & 5.3 & 0.3 & 3.9 & 0.6 & 4.6 & 0.6 & 9.1 & 0.3 & 2.6 & 0.5\\
7637.51 & [Fe II] & 1.9 & 0.5 & 2.5 & 1.0 & 4.3 & 0.7 & 5.6 & 0.3 & 1.4 & 1.2 \\
7665.28 & [Fe II] & ... &  & ... &  & ... &  & 1.0 & 0.3 & ... & \\
7686.93 & [Fe II] & 1.0 & 0.3 & 3.7 & 1.1 & 1.7 & 0.7 & 2.9 & 0.3 & ... & \\
7891.86 & [Fe XI] & 2.4 & 0.8 & ... &  & 3.8 & 0.9 & 5.5 & 0.6 & ... & \\
8125.5 & [Cr II] & ... &  & ... &  & ... &  & 2.4 & 0.4 & ... & \\
8229.8 & [Cr II] & ... &  & ... &  & ... &  & 1.1 & 0.2 & 0.5 & 0.3\\
8234.54 & [Fe IX] & 0.8 & 0.3 & ... &  & 2.5 & 0.4 & 3.7 & 0.3 & ... & \\
8542.1 & Ca II & 3.1 & 0.7 & ... &  & 3.3 & 0.7 & 3.7 & 0.6 & 4.2 & 1.0\\
8578.69 & [Cl II] & 0.5 & 0.3 & ... &  & ... &  & 2.3 & 0.5 & ... & \\
8616.95 & [Fe II] & 19.2 & 0.9 & 20.6 & 1.4 & 20.7 & 1.1 & 30 & 0.6 & 25.7 & 2.8\\
8662.14 & [Ca II] & 0.8 & 0.3 & ... &  & ... &  & ... &  & ... & \\
8727.13 & [C I] & ... &  & ... &  & ... &  & 1.3 & 0.2 & ... & \\
8891.93 & [Fe II] & 7.7 & 1.0 & 5.3 & 4.2 & 5.2 & 3.5 & 9.5 & 0.6 & ... & \\
 \hline
F(H$\beta$) $\times10^{16}$ & (erg/cm$^2$/s) & 86.17 &  & 45.67 &  & 49.32 &  & 106.87 &  & 38.71 & \\
\hline
 \end{tabular}}
\end{table*}

\begin{table*}
 \centering
 \small
   \caption{Measured line fluxes for N132D shocked cloudlets P06 -- P10}
    \label{tableA2}
   \scalebox{0.5}{
  \begin{tabular}{lccccccccccc}
 \hline
\bf{Cloud ID\#:} &  & \bf{P06} &  & \bf{P07} &  & \bf{P08} &  & \bf{P09} &  & \bf{P10} & \\
Lambda & ID & Flux & Error & Flux & Error & Flux & Error & Flux & Error & Flux & Error\\
 \hline
3736.02 & [O II] & 174.3 & 7.4 & 93.6 & 29.2 & 67.3 & 7.6 & 77.2 & 15.5 & 138.2 & 8.3 \\
3728.82 & [O II] & 61.8 & 5.9 & 54.9 & 28.8 & 33.1 & 5.3 & ... &  & 120.4 & 8.3  \\
3835.4 & H$\eta$ & 12.2 & 2.7 & 6.6 & 1.8 & 14.5 & 4.8 & 16.0 & 3.5 & ... & \\
3868.76 & [Ne III] & 16.5 & 3.1 & 19.7 & 2.0 & 12.7 & 9.3 & ... &  & 34.8 & 3.7 \\
3888.8 & HeI, H & 24.0 & 3.6 & 14.2 & 2.3 & 14.4 & 3.5 & ... &  & 6.6 & 2.8 \\
3933.66 & Ca II & 15.0 & 2.1 & 14.6 & 3.6 & 14.3 & 5.6 & 23.2 & 7.8 & ... & \\
3969 & [Ne III], Ca II, H$\epsilon$ & 30.5 & 3.3 & 21.8 & 2.0 & 37.3 & 5.4 & ... &  & 25.8 & 4.2 \\
4026.21 & He I & ... &  & 4.2 & 1.5 & ... &  & ... &  & 5.2 & 1.1 \\
4068.6 & [S II] & 33.4 & 1.7 & 28.2 & 1.1 & 15.6 & 3.3 & 21.9 & 2.6 & 25.0 & 1.6 \\
4076.35 & [S II] & 16.0 & 1.9 & 7.2 & 1.0 & 5.5 & 2.3 & 14.7 & 2.4 & 11.0 & 2.0 \\
4101.74 & H$\delta$ & 25.3 & 2.6 & 17.9 & 1.2 & 38.5 & 5.1 & 22.0 & 3.5 & 29.4 & 2.3 \\
4244 & [Fe II] & ... &  & 3.9 & 0.9 & ... &  & ... &  & 6.1 & 2.7 \\
4276.83 & [Fe II] & ... &  & 2.9 & 0.8 & ... &  & ... &  & ... & \\
4287.43 & [Fe II] & ... &  & 1.5 & 0.6 & ... &  & ... &  & ... & \\
4340.47 & H$\gamma$ & 46.0 & 1.7 & 51.9 & 2.6 & 53.1 & 3.1 & 38.7 & 2.0 & 41.5 & 1.3 \\
4359.14 & [Fe IX] & 8.3 & 5.6 & ... &  & 5.3 & 2.6 & ... &  & 2.2 & 0.9 \\
4363.21 & [O III] & 5.1 & 6.2 & 8.4 & 1.9 & 7.1 & 1.5 & ... &  & 11.0 & 2.0 \\
4416.27 & [Fe II] & 3.5 & 2.0 & 6.3 & 1.7 & ... &  & ... &  & ... & \\
4452.38 & [Fe II] & ... &  & 4.5 & 0.7 & ... &  & ... &  & 2.0 & 1.0 \\
4471.5 & He I & 1.1 & 0.4 & 1.9 & 0.5 & ... &  & ... &  & 3.6 & 1.1 \\
4658.5 & [O III],[Fe III] & 7.5 & 0.8 & 3.3 & 0.6 & 3.8 & 0.9 & 4.1 & 1.5 & 4.9 & 1.1 \\
4685.7 & He II & 8.2 & 1.1 & 6.2 & 0.7 & 1.4 & 0.6 & ... &  & 8.6 & 1.4 \\
4701.53 & [Fe III] & ... &  & 2.4 & 0.7 & 5.9 & 1.1 & ... &  & ... & \\
4754.68 & [Fe III] & ... &  & 4.6 & 1.9 & ... &  & ... &  & ... & \\
4814.54 & [Fe II] & 4.4 & 0.9 & 2.5 & 0.5 & ... &  & ... &  & ... & \\
4861.29 & H$\beta$ & 100.0 & 1.0 & 100.0 & 6.4 & 100.0 & 2.4 & 100.0 & 1.5 & 100.0 & 1.7 \\
4881 & [Fe III] & 4.0 & 0.7 & ... &  & ... &  & ... &  & 0.7 & 0.5 \\
4889.62 & [Fe II] & ... &  & ... &  & 2.5 & 1.1 & 2.3 & 0.8 & ... & \\
4905.35 & [Fe II] & 4.6 & 1.2 & ... &  & 4.2 & 2.9 & ... &  & ... & \\
4958.91 & [O III] & 15.0 & 1.3 & 31.9 & 5.1 & 14.5 & 14.7 & ... &  & 47.6 & 1.6 \\
5006.84 & [O III] & 61.2 & 1.1 & 115.3 & 6.3 & 43.3 & 1.4 & ... &  & 136.6 & 1.6 \\
5158.79 & [Fe II], [Fe VII] & 15.2 & 1.4 & 11.3 & 0.5 & 15.6 & 1.7 & 14.4 & 1.0 & 5.0 & 1.1 \\
5198.8 & [N I] & 2.2 & 1.2 & 0.4 & 0.2 & ... &  & ... &  & ... & \\
5261.66 & [Fe II] & 5.5 & 0.9 & 5.3 & 0.5 & 2.8 & 0.8 & 6.8 & 1.1 & ... & \\
5270.4 & [Fe III] & 3.3 & 2.0 & 1.0 & 0.5 & ... &  & ... &  & ... & \\
5273.36 & [Fe II] & ... &  & 0.5 & 0.2 & ... &  & ... &  & ... & \\
5303.06 & [Fe XIV] & 6.6 & 1.5 & 5.5 & 0.9 & 6.7 & 2.6 & 9.1 & 1.1 & 6.0 & 1.3 \\
5334.5 & [Fe II],[Fe VI] & ... &  & ... &  & 4.9 & 2.0 & 3.1 & 0.7 & ... & \\
5376 & [Fe II],[Fe VIII] & ... &  & 1.4 & 1.1 & ... &  & ... &  & 3.1 & 1.6 \\
5527.3 & [Fe II] & 1.6 & 2.1 & ... &  & ... &  & ... &  & ... & \\
5754.59 & [N II] & 2.9 & 1.2 & ... &  & ... &  & ... &  & ... & \\
5875.61 & He I & 9.4 & 1.1 & 5.2 & 2.5 & 6.1 & 3.3 & 12.3 & 1.0 & 5.9 & 1.4 \\
6086.97 & [Fe VII] & 1.4 & 0.4 & 1.4 & 0.5 & ... &  & 1.7 & 0.8 & 2.7 & 0.6 \\
6300.3 & [O I] & 86.7 & 5.8 & 58.8 & 0.9 & 57.3 & 9.4 & 96.7 & 1.4 & 71.3 & 1.0 \\
6312.06 & [S III] & 0.3 & 0.5 & ... &  & ... &  & 8.1 & 3.4 & ... & \\
6363.78 & [O I] & 29.1 & 2.1 & 20.8 & 0.9 & 20.4 & 8.9 & 30.6 & 0.8 & 24.8 & 0.8 \\
6374.52 & [Fe X] & 3.2 & 1.2 & ... &  & ... &  & ... &  & 4.2 & 0.7 \\
6548.05 & [N II] & 26.0 & 0.9 & 25.2 & 1.3 & 19.4 & 2.2 & 26.1 & 1.9 & 24.7 & 1.4 \\
6562.82 & H$\alpha$ & 319.0 & 1.4 & 298.7 & 2.7 & 298.2 & 4.1 & 300.0 & 2.5 & 300.6 & 1.7 \\
6583.45 & [N II] & 78.3 & 0.8 & 71.8 & 0.7 & 53.9 & 2.1 & 82.8 & 1.2 & 83.7 & 0.8 \\
6678.15 & He I & 2.0 & 0.5 & ... &  & 3.6 & 0.6 & 3.7 & 0.5 & 3.1 & 0.7 \\
6716.44 & [S II] & 42.0 & 0.6 & 59.8 & 0.4 & 57.1 & 1.0 & 80.2 & 1.0 & 43.8 & 0.6 \\
6730.82 & [S II] & 73.8 & 0.6 & 106.7 & 0.5 & 97.0 & 0.9 & 124.5 & 0.9 & 73.5 & 0.7 \\
7065.25 & He I & 4.6 & 0.5 & 5.2 & 0.5 & 7.6 & 1.2 & 9.6 & 1.5 & 4.1 & 0.6 \\
7135.79 & [Ar III] & 1.5 & 0.4 & 3.9 & 0.5 & 2.1 & 0.6 & 3.9 & 0.5 & 4.9 & 0.5 \\
7155.1 & [Fe II] & 22.4 & 0.6 & 15.9 & 0.5 & 19.4 & 0.9 & 21.4 & 0.6 & 12.0 & 0.6 \\
7172 & [Fe II] & 3.8 & 0.5 & 2.4 & 0.4 & 2.0 & 0.5 & 2.6 & 0.5 & 3.8 & 0.7 \\
7291.47 & [Ca II] & 39.9 & 0.9 & 35.0 & 2.7 & 44.2 & 2.1 & 45.8 & 1.1 & 12.6 & 1.2 \\
7321.5 & [Ca II], [O II] & 57.6 & 3.0 & 39.0 & 1.3 & ... &  & 47.3 & 2.6 & 46.6 & 4.0 \\
7330 & [O II] & 13.2 & 1.5 & 14.5 & 0.9 & ... &  & 4.2 & 1.7 & 35.9 & 2.4 \\
7377.82 & [Ni II] & 12.7 & 0.7 & 5.9 & 0.7 & 11.8 & 1.4 & 10.8 & 1.3 & 5.6 & 1.0 \\
7388.2 & [Fe II] & 6.5 & 1.3 & 4.2 & 7.7 & 4.1 & 1.3 & ... &  & ... & \\
7411.61 & [Ni II] & 1.7 & 0.4 & ... &  & ... &  & 1.7 & 0.5 & ... & \\
7452.5 & [Fe II] & 9.6 & 0.9 & 5.6 & 0.7 & 7.1 & 1.1 & 4.5 & 0.6 & 3.1 & 0.5 \\
7637.51 & [Fe II] & 4.0 & 0.6 & 2.0 & 0.8 & 7.2 & 1.5 & 5.4 & 1.1 & ... & \\
7665.28 & [Fe II] & ... &  & ... &  & ... &  & ... &  & ... & \\
7686.93 & [Fe II] & 1.7 & 0.6 & 1.0 & 0.4 & ... &  & 1.3 & 0.4 & 1.3 & 0.5 \\
7891.86 & [FeXI] & 1.9 & 0.7 & ... &  & ... &  & 5.7 & 1.4 & ... & \\
8125.5 & [Cr II] & ... &  & 1.5 & 0.4 & 3.6 & 0.9 & 2.8 & 0.9 & ... & \\
8229.8 & [Cr II] & 1.2 & 0.2 & ... &  & 0.8 & 0.4 & 3.0 & 0.8 & ... & \\
8234.54 & [Fe IX] & 3.6 & 0.4 & 2.9 & 0.8 & ... &  & 2.7 & 0.6 & ... & \\
8542.1 & Ca II & ... &  & 4.9 & 1.0 & ... &  & 1.4 & 0.4 & ... & \\
8578.69 & [Cl II] & ... &  & 1.9 & 0.3 & ... &  & ... &  & ... & \\
8616.95 & [Fe II] & 29.0 & 1.4 & 19.0 & 0.6 & 19.1 & 2.7 & 26.1 & 1.4 & 11.0 & 1.4 \\
8662.14 & [Ca II] & ... &  & 1.8 & 0.6 & ... &  & 1.4 & 0.5 & ... & \\
8727.13 & [C I] & 1.7 & 0.6 & 1.3 & 0.4 & ... &  & ... &  & 5.7 & 1.6 \\
8862.78 & H I & ... &  & ... &  & ... &  & 5.4 & 0.9 & ... & \\
8891.93 & [Fe II] & 5.5 & 1.5 & 4.5 & 3.5 & ... &  & 6.9 & 3.0 & 4.2 & 1.1 \\
  \hline
F(H$\beta$) $\times10^{16}$ &(erg/cm$^2$/s) & 48.9 &  & 76.5 &  & 76.5 &  & 56.8 &  & 54.8 & \\
\hline
 \end{tabular}}
\end{table*}

\begin{table*}
 \centering
 \small
   \caption{Measured line fluxes for N132D shocked cloudlets P11 -- P15}
    \label{tableA3}
   \scalebox{0.5}{
  \begin{tabular}{lccccccccccc}
 \hline
\bf{Cloud ID\#:} &  & \bf{P11} &  & \bf{P12} &  & \bf{P13} &  & \bf{P14} &  & \bf{P15} & \\
Lambda & ID & Flux & Error & Flux & Error & Flux & Error & Flux & Error & Flux & Error\\
 \hline
3736.0 & [O II] & 88.5 & 17.4 & 216.6 & 4.5 & 223.3 & 12.4 & 229.5 & 2.0 & 295.7 & 6.7 \\
3728.8 & [O II] & 197.4 & 24.0 & 106.7 & 4.4 & 148.8 & 12.4 & 119.5 & 2.0 & 208.4 & 6.7 \\
3835.4 & H$\eta$ & ... &  & 6.5 & 0.7 & ... &  & 5.5 & 0.5 & 8.2 & 0.9 \\
3868.8 & [Ne III] & 41.7 & 4.1 & 29.9 & 0.5 & 23.3 & 2.8 & 32.2 & 0.5 & 35.6 & 0.8 \\
3888.8 & HeI, H & 14.2 & 5.7 & 15.9 & 0.7 & 11.5 & 2.8 & 19.8 & 0.4 & 21.8 & 0.8 \\
3933.7 & Ca II & ... &  & 11.8 & 0.4 & ... &  & 16.8 & 0.4 & 13.9 & 1.1 \\
3969.0 & [Ne III], Ca II, H$\epsilon$ & 41.6 & 10.9 & 30.8 & 0.7 & ... &  & 32.5 & 0.6 & 34.8 & 1.2 \\
4026.2 & He I & 11.7 & 3.5 & 2.0 & 0.5 & ... &  & 1.6 & 0.3 & 1.2 & 0.4 \\
4068.6 & [S II] & 31.7 & 3.7 & 27.5 & 0.4 & 22.8 & 2.2 & 41.4 & 0.3 & 29.2 & 0.7 \\
4076.4 & [S II] & 15.4 & 2.4 & 9.2 & 0.3 & 3.0 & 1.2 & 14.1 & 0.3 & 9.7 & 0.6 \\
4101.7 & H$\delta$ & 23.0 & 3.6 & 22.2 & 0.5 & 12.2 & 2.5 & 24.5 & 0.3 & 28.0 & 0.6 \\
4244.0 & [Fe II] & ... &  & 4.3 & 0.2 & ... &  & 4.9 & 0.2 & 3.8 & 0.5 \\
4276.8 & [Fe II] & 8.5 & 2.4 & 0.9 & 0.2 & ... &  & 1.5 & 0.2 & 2.4 & 0.7 \\
4287.4 & [Fe II] & ... &  & 3.0 & 0.2 & 2.7 & 1.5 & 3.9 & 0.2 & 3.6 & 0.5 \\
4319.6 & [Fe II] & ... &  & 0.5 & 0.1 & ... &  & 0.7 & 0.2 & ... &  \\
4340.5 & H$\gamma$ & 53.6 & 2.4 & 42.0 & 0.6 & 38.9 & 1.5 & 44.8 & 0.4 & 48.7 & 0.6 \\
4359.1 & [Fe IX] & ... &  & 2.7 & 0.5 & ... &  & 4.1 & 0.4 & 3.7 & 0.6 \\
4363.2 & [O III] & 14.3 & 1.9 & 7.8 & 0.4 & 7.8 & 1.4 & 7.1 & 0.3 & 7.2 & 0.5 \\
4416.3 & [Fe II] & ... &  & 4.7 & 0.4 & ... &  & 4.9 & 0.3 & 7.2 & 0.6 \\
4452.4 & [Fe II] & ... &  & 0.6 & 0.1 & ... &  & 1.0 & 0.1 & 1.4 & 0.3 \\
4458.0 & [Fe II] & ... &  & 1.1 & 0.2 & ... &  & 0.9 & 0.2 & 1.2 & 0.5 \\
4471.5 & He I & 3.6 & 1.2 & 3.8 & 0.3 & 5.2 & 1.6 & 3.9 & 0.2 & 4.5 & 0.4 \\
4566.8 & Mg I] & ... & ... & 2.5 & 0.5 & ... & ... & 3.6 & 0.2 & 1.8 & 0.4 \\
4658.5 & [O III],[Fe III] & 2.9 & 1.0 & 5.0 & 0.2 & 3.4 & 0.8 & 5.1 & 0.1 & 7.0 & 0.3 \\
4685.7 & He II & ... &  & 6.5 & 0.2 & 3.8 & 1.2 & 5.5 & 0.2 & 5.9 & 0.3 \\
4701.5 & [Fe III] & ... &  & 4.1 & 1.3 & ... &  & 1.1 & 0.1 & 1.9 & 0.3 \\
4754.7 & [Fe III] & ... &  & 0.7 & 0.2 & ... &  & 0.6 & 0.2 & 2.0 & 0.3 \\
4814.5 & [Fe II] & ... &  & 2.3 & 0.2 & 5.1 & 2.3 & 2.4 & 0.2 & 2.1 & 0.2 \\
4861.3 & H$\beta$ & 100.0 & 3.1 & 100.0 & 1.2 & 100.0 & 1.1 & 100.0 & 0.3 & 100.0 & 0.7 \\
4881.0 & [Fe III] & ... &  & 2.2 & 0.9 & ... &  & 2.1 & 0.2 & 2.9 & 0.4 \\
4889.6 & [Fe II] & ... &  & 2.0 & 0.9 & ... &  & 2.2 & 0.2 & 1.7 & 0.4 \\
4905.4 & [Fe II] & ... &  & 0.9 & 0.9 & ... &  & 0.9 & 0.2 & 0.6 & 0.2 \\
4958.9 & [O III] & 76.7 & 19.7 & 39.3 & 1.2 & 35.5 & 1.1 & 38.4 & 0.3 & 33.9 & 0.6 \\
5006.8 & [O III] & 249.2 & 4.5 & 120.5 & 1.5 & 116.1 & 1.4 & 114.9 & 0.8 & 103.6 & 1.3 \\
5111.6 & [Fe II] & ... &  & 2.8 & 1.0 & ... &  & 1.6 & 0.1 & 1.0 & 0.2 \\
5158.8 & [Fe II], [Fe VII] & 7.2 & 1.8 & 11.3 & 0.5 & 14.9 & 1.1 & 11.7 & 0.2 & 10.7 & 0.2 \\
5198.8 & [N I] & 1.4 & 0.6 & 4.2 & 0.7 & 18.1 & 2.5 & 3.1 & 0.3 & 2.9 & 0.4 \\
5261.7 & [Fe II] & 4.5 & 1.3 & 4.1 & 0.4 & 2.4 & 0.7 & 4.2 & 0.1 & 3.8 & 0.3 \\
5270.4 & [Fe III] & 6.1 & 1.2 & 3.7 & 0.9 & ... &  & 0.9 & 1.3 & ... &  \\
5273.4 & [Fe II] & 3.4 & 1.4 & 1.2 & 0.8 & ... &  & 3.3 & 1.4 & 4.6 & 0.4 \\
5303.1 & [Fe XIV] & 30.4 & 1.9 & 3.2 & 0.6 & 4.3 & 1.5 & 3.7 & 0.2 & 1.6 & 0.3 \\
5334.5 & [Fe II],[Fe VI] & 5.1 & 1.8 & 2.0 & 0.5 & 1.5 & 0.7 & 2.2 & 0.1 & 1.7 & 0.2 \\
5376.0 & [Fe II],[Fe VIII] & ... &  & 0.5 & 0.2 & ... &  & 1.1 & 0.2 & 0.2 & 0.1 \\
5527.3 & [Fe II] & ... &  & 2.5 & 0.3 & ... &  & 2.6 & 0.3 & 1.9 & 0.3 \\
5754.6 & [N II] & ... &  & 1.2 & 0.3 & 4.1 & 0.9 & 1.3 & 0.1 & 1.8 & 0.2 \\
5875.6 & He I & 10.2 & 1.9 & 8.0 & 0.4 & 9.1 & 0.8 & 11.5 & 0.1 & 10.4 & 0.3 \\
6087.0 & [Fe VII] & 1.3 & 0.7 & 1.2 & 0.5 & ... &  & 0.6 & 0.1 & 1.6 & 0.2 \\
6300.3 & [O I] & 43.5 & 1.0 & 49.5 & 1.5 & 65.5 & 3.3 & 104.5 & 1.3 & 79.0 & 0.5 \\
6312.1 & [S III] & ... &  & ... &  & 1.4 & 1.0 & 0.9 & 0.4 & 1.2 & 0.7 \\
6363.8 & [O I] & 16.3 & 0.8 & 12.4 & 1.2 & 22.6 & 1.3 & 35.2 & 0.3 & 26.8 & 0.4 \\
6374.5 & [Fe X] & 13.3 & 1.0 & ... &  & 6.0 & 2.2 & 2.7 & 0.3 & 3.7 & 0.3 \\
6548.1 & [N II] & 18.8 & 1.9 & 24.3 & 1.0 & 63.3 & 0.6 & 29.5 & 1.0 & 30.2 & 0.5 \\
6562.8 & H$\alpha$ & 296.9 & 3.0 & 298.0 & 2.1 & 295.6 & 1.7 & 305.4 & 1.3 & 297.6 & 0.9 \\
6583.5 & [N II] & 52.9 & 1.2 & 70.8 & 0.6 & 197.1 & 1.1 & 91.4 & 0.4 & 93.6 & 0.6 \\
6678.2 & He I & 2.7 & 0.4 & 1.8 & 0.3 & 1.9 & 0.4 & 3.2 & 0.1 & 2.7 & 0.1 \\
6716.4 & [S II] & 31.2 & 0.8 & 61.8 & 0.5 & 82.3 & 0.7 & 85.8 & 0.3 & 94.3 & 0.3 \\
6730.8 & [S II] & 61.4 & 0.8 & 107.5 & 0.5 & 119.5 & 0.9 & 144.9 & 0.4 & 134.9 & 0.4 \\
7065.3 & He I & 4.1 & 0.7 & 3.3 & 0.3 & 4.0 & 0.4 & 5.2 & 0.1 & 4.0 & 0.2 \\
7135.8 & [Ar III] & 6.8 & 0.7 & 3.1 & 0.2 & 3.8 & 0.4 & 4.6 & 0.1 & 4.2 & 0.2 \\
7155.1 & [Fe II] & 14.8 & 0.7 & 13.6 & 0.3 & 13.8 & 0.5 & 22.1 & 0.1 & 16.3 & 0.2 \\
7172.0 & [Fe II] & 2.3 & 0.4 & 3.4 & 0.3 & 2.8 & 0.5 & 4.7 & 0.1 & 2.3 & 0.2 \\
7291.5 & [Ca II] & 21.0 & 0.9 & 30.8 & 3.1 & 26.3 & 0.5 & 39.1 & 0.3 & 26.9 & 0.4 \\
7321.5 & [Ca II], [O II] & 35.5 & 1.5 & 38.2 & 1.5 & 32.3 & 1.5 & 51.9 & 0.8 & 36.0 & 0.7 \\
7330.0 & [O II] & 23.9 & 1.2 & 14.5 & 1.1 & 11.5 & 0.9 & 18.8 & 0.5 & 14.1 & 0.5 \\
7377.8 & [Ni II] & 6.8 & 0.9 & 5.3 & 0.5 & 6.6 & 0.7 & 10.2 & 0.1 & 8.8 & 0.3 \\
7388.2 & [Fe II] & 1.8 & 0.5 & 1.7 & 0.4 & ... &  & 3.2 & 0.1 & 2.2 & 0.2 \\
7411.6 & [Ni II] & 53.8 & 25.0 & ... &  & 0.8 & 0.2 & 0.6 & 0.1 & 0.8 & 0.2 \\
7452.5 & [Fe II] & 5.9 & 0.8 & 4.9 & 0.6 & 6.7 & 0.5 & 7.0 & 0.1 & 5.6 & 0.2 \\
7637.5 & [Fe II] & ... &  & 1.4 & 0.3 & 0.9 & 0.4 & 4.3 & 0.1 & 2.7 & 0.3 \\
7665.3 & [Fe II] & ... &  & ... &  & 0.5 & 0.3 & 0.8 & 0.1 & 0.5 & 0.2 \\
7686.9 & [Fe II] & ... &  & 1.2 & 0.3 & 1.3 & 0.5 & 2.0 & 0.1 & 1.5 & 0.2 \\
7891.9 & [Fe XI] & 7.9 & 1.4 & ... &  & ... &  & 2.8 & 0.3 & 2.8 & 0.4 \\
8125.5 & [Cr II] & ... &  & 0.8 & 0.4 & 2.0 & 0.7 & 1.6 & 0.1 & 1.3 & 0.2 \\
8229.8 & [Cr II] & ... &  & 3.8 & 1.4 & 3.0 & 1.3 & 1.3 & 0.1 & ... &  \\
8234.5 & [Fe IX] & 7.6 & 1.9 & ... &  & 1.3 & 0.5 & 2.0 & 0.1 & 2.1 & 0.4 \\
8542.1 & Ca II & ... &  & ... &  &... &  & 2.1 & 0.1 & ... &  \\
8578.7 & [Cl II] & ... &  & 2.6 & 1.0 & 33.2 & 14.9 & 1.8 & 0.1 & ... &  \\
8617.0 & [Fe II] & 24.4 & 1.5 & 17.3 & 1.2 & 18.0 & 1.0 & 27.9 & 0.3 & 18.8 & 0.4 \\
8662.1 & [Ca II] & ... &  & ... &  & ... &  & 2.5 & 0.4 & 1.6 & 0.5 \\
8727.1 & [C I] & ... &  & 0.4 & 0.2 & ... &  & 1.2 & 0.1 & 1.2 & 0.1 \\
8862.8 & H I & ... &  & ... &  & ... &  & 0.3 & 0.2 & ... & \\
8891.9 & [Fe II] & 9.8 & 1.0 & 4.1 & 1.4 & 5.5 & 0.7 & 7.5 & 0.4 & 4.9 & 0.8 \\
  \hline
F(H$\beta$) $\times10^{16}$ &(erg/cm$^2$/s) & 33.3 &  & 353.5 &  & 61.8 &  & 576.6 &  & 254.2 & \\
\hline
 \end{tabular}}
\end{table*}

\begin{table*}
 \centering
 \small
   \caption{Measured line fluxes for N132D shocked cloudlets P16 -- P20}
    \label{tableA4}
   \scalebox{0.5}{
  \begin{tabular}{lccccccccccc}
 \hline
\bf{Cloud ID\#:} &  & \bf{P16} &  & \bf{P17} &  & \bf{P18} &  & \bf{P19} &  & \bf{P20} & \\
Lambda & ID & Flux & Error & Flux & Error & Flux & Error & Flux & Error & Flux & Error\\
 \hline
3736.02 & [O II] & 212.1 & 9.0 & 99.2 & 19.1 & 423.1 & 2.8 & 403.7 & 5.2 & 273.2 & 19.1 \\
3728.82 & [O II] & 120.2 & 10.3 & 411.7 & 22.5 & 91.9 & 2.2 & 105.0 & 3.9 & 146.1 & 13.0 \\
3835.40 & H$\eta$ & 63.7 & 10.2 & 10.2 & 1.9 & 9.3 & 0.7 & 3.3 & 2.0 & ... & \\ 
3868.76 & [Ne III] & 48.2 & 2.9 & 21.4 & 2.8 & 39.6 & 0.8 & 40.5 & 1.6 & 82.2 & 8.8 \\
3888.80 & HeI, H & 58.0 & 6.4 & 17.2 & 3.7 & 22.4 & 0.6 & 18.0 & 1.1 & 33.7 & 12.5 \\
3933.66 & Ca II & 19.1 & 5.1 & ... &  & 14.9 & 0.4 & 16.3 & 1.9 & ... &  \\
3969.00 & [Ne III], Ca II, H$\epsilon$ & 79.6 & 6.3 & 40.0 & 5.1 & 37.5 & 0.6 & 28.0 & 1.7 & ... &  \\
4026.21 & He I & ... &  & ... &  & 1.0 & 0.2 & 8.8 & 2.0 & ... &  \\
4068.60 & [S II] & 40.1 & 1.4 & 32.7 & 1.6 & 38.9 & 0.4 & 36.9 & 1.3 & 9.3 & 4.4 \\
4076.35 & [S II] & 15.2 & 1.2 & 14.1 & 2.8 & 13.9 & 0.4 & 10.7 & 1.0 & ... &  \\
4101.74 & H$\delta$ & 54.9 & 4.3 & 39.4 & 3.0 & 26.8 & 0.5 & 23.5 & 1.2 & 14.3 & 6.4 \\
4244.00 & [Fe II] & ... &  & ... &  & 4.9 & 0.3 & 6.1 & 0.6 & ... &  \\
4276.83 & [Fe II] & ... &  & ... &  & 0.8 & 0.2 & ... &  & ... &  \\
4287.43 & [Fe II] & ... &  & ... &  & 4.0 & 0.3 & 4.0 & 0.8 & ... & \\ 
4319.60 & [Fe II] & ... &  & ... &  & 0.4 & 0.3 & ... &  & ... & \\ 
4340.47 & H$\gamma$ & 58.2 & 3.2 & 53.1 & 2.2 & 47.4 & 0.8 & 41.5 & 1.8 & 44.8 & 5.1 \\
4359.14 & [Fe IX] & ... &  & ... &  & 8.0 & 1.5 & 4.8 & 1.4 & ... &  \\
4363.21 & [O III] & 19.6 & 4.5 & 10.7 & 1.5 & 4.7 & 0.4 & 6.6 & 1.2 & 84.2 & 3.6 \\
4416.27 & [Fe II] & ... &  & 2.1 & 0.9 & 5.6 & 0.4 & 3.3 & 0.5 & ... & \\ 
4452.38 & [Fe II] & ... &  & ... &  & 1.6 & 0.4 & 1.6 & 0.7 & ... &  \\
4457.95 & [Fe II] & ... &  & ... &  & 1.1 & 0.2 & 0.9 & 0.3 & ... &  \\
4471.50 & He I & 1.3 & 0.5 & ... &  & 5.6 & 0.4 & 4.7 & 0.4 & ... &  \\
4658.50 & [O III],[Fe III] & 4.3 & 0.7 & 4.9 & 1.4 & 5.9 & 0.2 & 6.3 & 0.4 & 11.0 & 2.3 \\
4685.70 & He II & 7.7 & 1.0 & 3.1 & 1.0 & 6.1 & 0.2 & 5.5 & 0.5 & 12.9 & 3.1 \\
4701.53 & [Fe III] & ... &  & ... &  & 1.7 & 0.3 & 1.2 & 0.3 & 6.9 & 2.8 \\
4754.68 & [Fe III] & 2.4 & 0.9 & ... &  & 1.2 & 0.2 & 1.4 & 0.4 & ... &  \\
4814.54 & [Fe II] & ... &  & 12.7 & 4.3 & 2.0 & 0.2 & 1.8 & 0.4 & ... &  \\
4861.29 & H$\beta$ & 100.0 & 1.8 & 100.0 & 1.6 & 100.0 & 1.7 & 100.0 & 2.9 & 100.0 & 3.4 \\
4881.00 & [Fe III] & 2.6 & 0.9 & ... &  & 2.8 & 0.7 & 1.5 & 0.7 & ... &  \\
4889.62 & [Fe II] & ... &  & ... &  & 2.2 & 0.6 & ... &  & ... &  \\
4905.35 & [Fe II] & ... &  & ... &  & 1.2 & 0.8 & ... &  & ... &  \\
4958.91 & [O III] & 85.8 & 1.2 & 38.8 & 1.4 & 40.4 & 1.3 & 40.3 & 1.0 & 326.2 & 4.0 \\
5006.84 & [O III] & 246.5 & 2.6 & 134.8 & 2.1 & 123.4 & 2.9 & 130.0 & 1.8 & 967.4 & 6.6 \\
5111.64 & [Fe II] & ... &  & ... &  & 1.3 & 0.2 & 0.9 & 0.4 & ... &  \\
5158.79 & [Fe II], [Fe VII] & 5.4 & 0.7 & 7.4 & 1.4 & 12.1 & 0.3 & 9.2 & 0.5 & 11.3 & 2.7 \\
5198.80 & [N I] & ... &  & 3.8 & 1.5 & 3.7 & 0.2 & 3.7 & 0.9 & ... &  \\
5261.66 & [Fe II] & ... &  & 2.2 & 0.6 & 4.2 & 0.2 & 3.1 & 0.4 & ... &  \\
5270.40 & [Fe III] & ... &  & ... &  & 4.4 & 0.6 & 2.3 & 0.8 & ... &  \\
5273.36 & [Fe II] & ... &  & ... &  & 0.4 & 0.6 & ... &  & ... &  \\
5303.06 & [Fe XIV] & 3.2 & 1.3 & ... &  & 1.8 & 0.2 & ... &  & 4.2 & 1.7 \\
5334.50 & [Fe II],[Fe VI] & 3.9 & 1.3 & ... &  & 2.2 & 0.2 & 2.0 & 0.3 & ... &  \\
5376.00 & [Fe II],[Fe VIII] & 1.1 & 0.5 & ... &  & 0.7 & 0.1 & 2.8 & 5.1 & ... &  \\
5527.30 & [Fe II] & ... &  & ... &  & 1.9 & 0.6 & 3.2 & 0.9 & ... &  \\
5754.59 & [N II] & ... &  & 11.4 & 4.0 & 2.2 & 0.1 & 2.5 & 0.5 & 8.0 & 2.2 \\
5875.61 & He I & 7.0 & 0.7 & 7.9 & 1.7 & 10.9 & 0.2 & 11.5 & 0.5 & 6.0 & 3.5 \\
6086.97 & [Fe VII] & 1.8 & 0.6 & ... &  & 0.8 & 0.1 & ... &  & ... &  \\
6300.30 & [O I] & 78.1 & 2.8 & 97.4 & 3.9 & 96.4 & 0.6 & 74.4 & 1.0 & ... &  \\
6312.06 & [S III] & 3.1 & 1.8 & ... &  & 2.2 & 0.9 & ... &  & ... &  \\
6363.78 & [O I] & 20.0 & 0.8 & 20.5 & 3.9 & 31.6 & 0.3 & 27.5 & 0.6 & ... &  \\
6374.52 & [Fe X] & 3.2 & 0.9 & ... &  & 2.3 & 0.4 & ... &  & ... &  \\
6548.05 & [N II] & 29.3 & 1.2 & 36.1 & 1.4 & 31.6 & 2.6 & 32.5 & 3.3 & 33.4 & 2.4 \\
6562.82 & H$\alpha$ & 298.0 & 2.1 & 300.9 & 3.3 & 298.9 & 3.0 & 300.7 & 6.0 & 294.6 & 3.9 \\
6583.45 & [N II] & 86.3 & 1.0 & 114.1 & 1.5 & 98.4 & 1.6 & 99.0 & 3.0 & 95.2 & 1.6 \\
6678.15 & He I & 2.8 & 0.5 & 2.5 & 0.4 & 2.9 & 0.1 & 3.4 & 0.3 & ... &  \\
6716.44 & [S II] & 61.4 & 0.6 & 113.9 & 1.4 & 87.5 & 1.9 & 92.3 & 2.9 & 121.2 & 2.7 \\
6730.82 & [S II] & 109.6 & 0.6 & 165.1 & 2.0 & 132.8 & 2.0 & 145.6 & 4.1 & 141.6 & 3.0 \\
7065.25 & He I & 4.1 & 0.5 & 0.8 & 0.4 & 4.4 & 0.1 & 3.9 & 0.2 & ... &  \\
7135.79 & [Ar III] & 9.2 & 0.8 & 3.8 & 0.6 & 5.1 & 0.2 & 3.8 & 0.2 & 16.6 & 1.6 \\
7155.10 & [Fe II] & 9.1 & 0.5 & 2.0 & 0.6 & 18.0 & 0.2 & 17.7 & 0.5 & 2.6 & 0.8 \\
7172.00 & [Fe II] & 2.1 & 0.6 & 0.9 & 0.3 & 3.7 & 0.2 & 3.7 & 0.2 & ... &  \\
7291.47 & [Ca II] & 7.4 & 0.8 & 0.8 & 0.4 & 26.9 & 0.3 & 26.4 & 0.6 & ... &  \\
7321.50 & [Ca II], [O II] & 37.7 & 1.2 & 28.2 & 1.4 & 42.1 & 0.4 & 40.9 & 1.6 & 45.9 & 4.4 \\
7330.00 & [O II] & 27.5 & 1.4 & 19.6 & 1.0 & 19.4 & 0.3 & 12.6 & 0.9 & 24.8 & 2.6 \\
7377.82 & [Ni II] & 3.2 & 0.4 & ... &  & 8.2 & 0.2 & 7.7 & 0.3 & 6.9 & 2.0 \\
7388.20 & [Fe II] & 1.7 & 0.8 & ... &  & 2.8 & 0.2 & 2.0 & 0.3 & ... &  \\
7411.61 & [Ni II] & ... &  & ... &  & 0.8 & 0.1 & ... &  & ... &  \\
7452.50 & [Fe II] & 3.1 & 0.6 & ... &  & 5.7 & 0.1 & 5.4 & 0.3 & ... &  \\
7637.51 & [Fe II] & 3.0 & 1.6 & ... &  & 2.9 & 0.1 & 2.3 & 0.4 & ... &  \\
7665.28 & [Fe II] & ... &  & ... &  & 0.6 & 0.1 & ... &  & ... &  \\
7686.93 & [Fe II] & 1.2 & 0.5 & ... &  & 1.7 & 0.1 & 1.6 & 0.2 & ... &  \\
7891.86 & [Fe XI] & ... &  & ... &  & 1.6 & 0.2 & 3.5 & 1.0 & ... &  \\
8125.50 & [Cr II] & ... &  & ... &  & 1.1 & 0.1 & 1.4 & 0.3 & ... &  \\
8229.80 & [Cr II] & ... &  & ... &  & 0.6 & 0.1 & ... &  & ... &  \\
8234.54 & [Fe IX] & ... &  & ... &  & 1.3 & 0.1 & 0.5 & 0.1 & ... &  \\
8542.10 & Ca II & 0.3 & 0.3 & ... &  & 1.8 & 0.2 & ... &  & ... &  \\
8578.69 & [Cl II] & ... &  & ... &  & 2.2 & 0.3 & ... &  & ... &  \\
8616.95 & [Fe II] & 12.9 & 1.7 & ... &  & 23.4 & 0.3 & ... &  & 12.6 & 2.9 \\
8662.14 & [Ca II] & 2.1 & 0.8 & ... &  & 2.4 & 0.4 & ... &  & ... &  \\
8727.13 & [C I] & 2.9 & 0.7 & ... &  & 1.0 & 0.1 & 1.2 & 0.2 & ... &  \\
8862.78 & H I & ... &  & ... &  & 0.5 & 0.2 & ... &  & ... &  \\
8891.93 & [Fe II] & 3.4 & 2.2 & ... &  & 5.4 & 0.6 & 4.9 & 0.4 & ... &  \\
  \hline
F(H$\beta$) $\times10^{16}$ &(erg/cm$^2$/s) & 53.6 &  & 47.7 &  & 544.3 &  &116.8 &  & 23.5 & \\
\hline
 \end{tabular}}
\end{table*}

\end{document}